\documentclass[superscriptaddress,showpacs,amssymb,10pt,aps,prd,print,longbibliography]{revtex4-1}
\usepackage{graphicx,epsfig,amssymb,times} 
\usepackage{amsmath,amsfonts}
\usepackage{bm}
\usepackage{epstopdf}
\usepackage[linktocpage,colorlinks]{hyperref}
\usepackage[caption=false]{subfig}
\usepackage[usenames]{color}  

\definecolor{coolblack}{rgb}{0.0, 0.18, 0.39}
\definecolor{darkred}{rgb}{0.5,0,0}
\definecolor{darkgreen}{rgb}{0,0.5,0}
\definecolor{darkblue}{rgb}{0,0,0.5}
\definecolor{lapislazuli}{rgb}{0.15, 0.38, 0.61}
\definecolor{venetianred}{rgb}{0.78, 0.03, 0.08}
\definecolor{bleudefrance}{rgb}{0.19, 0.55, 0.91}
\definecolor{dogwoodrose}{rgb}{0.84, 0.09, 0.41}
\hypersetup{colorlinks=true, citecolor=darkblue, linkcolor=darkblue, 
urlcolor = darkblue}

\def\be{\begin{equation}}
\def\ee{\end{equation}}

\newcommand{\bea}{\begin{eqnarray}}
\newcommand{\eea}{\end{eqnarray}}
\newcommand{\ben}{\begin{enumerate}}
\newcommand{\een}{\end{enumerate}}
\newcommand{\bi}{\begin{itemize}}
\newcommand{\ei}{\end{itemize}}

\newcommand{\nn}{\nonumber}

\renewcommand{\d}{\partial}

\def\ga{\mathrel{\raise.3ex\hbox{$>$\kern-.75em\lower1ex\hbox{$\sim$}}}}
\def\la{\mathrel{\raise.3ex\hbox{$<$\kern-.75em\lower1ex\hbox{$\sim$}}}}

\def\l{\left}
\def\r{\right}
\def\be{\begin{equation}}
\def\ee{\end{equation}}

\renewcommand{\d}{\rm{d}}

\def\I_M{{I_{\scriptscriptstyle M\times M}}}

\def\be{\begin{equation}}
\def\ee{\end{equation}}
\def\bea{\begin{eqnarray}}
\def\eea{\end{eqnarray}}
\newcommand{\beq}{\begin{eqnarray}}
\newcommand{\eeq}{\end{eqnarray}}

\newcommand{\beqal}{\begin{eqnarray}\label}
\newcommand{\beqa}{\begin{eqnarray}}
\newcommand{\eeqa}{\end{eqnarray}}

\newcommand {\non}{\nonumber\\}
\newcommand {\f}{\frac} 
\newcommand {\swf}{\mathsf{S}} 
\newcommand {\rtf}{\mathsf{R}} 
\newcommand {\swlm}{\mathfrak{s}\omega lm}

\newcommand {\sw}{\mathfrak{s}}

\newcommand{\omtil}{\widetilde{\omega}}

\begin{document}
\title{\large Scattering of massless bosonic fields by Kerr black holes: On-axis incidence}

\author{Luiz C. S. Leite}\email{luizcsleite@ufpa.br}
\affiliation{Faculdade de F\'{\i}sica, Universidade 
Federal do Par\'a, 66075-110, Bel\'em, Par\'a, Brazil.}

\author{Sam R.~Dolan}\email{s.dolan@sheffield.ac.uk}
\affiliation{Consortium for Fundamental Physics, School of 
Mathematics and Statistics, University of Sheffield, Hicks Building,
Hounsfield Road, Sheffield S3 7RH, United Kingdom}

\author{Lu\'is C. B. Crispino}\email{crispino@ufpa.br}
\affiliation{Faculdade de F\'{\i}sica, Universidade 
Federal do Par\'a, 66075-110, Bel\'em, Par\'a, Brazil.}
\begin{abstract}
We study the scattering of monochromatic bosonic plane 
waves impinging upon a rotating black hole, in the special case that the direction of incidence is aligned with the spin axis. We present accurate numerical results for electromagnetic Kerr scattering cross sections for the first time, and give a unified picture of the Kerr scattering for all massless bosonic fields.
\end{abstract}

\maketitle

\section{Introduction}
 
Scattering is a ubiquitous phenomenon in physics across all scales, from particle physics to the cosmic microwave background. The recent observation of ``chirps'' from binary mergers \cite{Abbott:2016blz} has shown beyond a reasonable doubt that black holes~(BHs) are abundant \cite{LIGOScientific:2018mvr} and that gravitational waves propagate at the speed of light \cite{TheLIGOScientific:2017qsa}. Of foundational interest, therefore, is the scattering of fundamental fields in the strongly curved spacetime geometry surrounding a BH. 

The time-independent scattering of fields by BHs has been studied since the late 1960s, with pioneering early contributions from Matzner and co-workers \cite{Matzner:1968, Matzner:1985rjn, Futterman:1988ni}, Mashhoon \cite{Mashhoon:1974cq}, and Sanchez \cite{Sanchez:1977vz}. Recent years have seen advances in calculating accurate scattering cross sections for rotating (Kerr) BHs, overcoming technical difficulties associated with the convergence of the partial-wave series. The massless scalar field ($s=0$) case was addressed in Ref.~\cite{Glampedakis:2001cx}, and the gravitational wave case ($s=2$) was given in Ref.~\cite{Dolan:2008kf}. 

The purpose of this paper is twofold: to present accurate numerical results for the scattering of the electromagnetic field ($s=1$) by a rotating BH for the first time, and to give a unified description of scattering for all massless boson fields ($s = 0, 1$ and $2$), complementing Refs.~\cite{Crispino:2009xt} and~\cite{PhysRevD.92.084056}, for Schwarzschild and Reissner-Nordstr\"om scattering, respectively, and Refs.~\cite{PhysRevD.84.084048} and~\cite{Leite:2017zyb}, for Reissner-Nordstr\"om and Kerr absorption, respectively.

We consider an idealized scenario, in which a planar wave of frequency $\omega$ impinges upon a Kerr BH of mass $M$ and angular momentum $J=aM$ along a direction parallel to its symmetry axis. This scenario is characterized by a pair of dimensionless parameters, $M\omega \equiv G M \omega / c^3$ and $a/M$. At low frequency $M\omega \ll1$, the scattering cross section is \cite{Dolan:2007ut}
\begin{align}
\lim_{M\omega \rightarrow 0} \left( \frac{1}{M^2} \frac{d\sigma}{d\Omega} \right) = \frac{\cos^{4s}(\theta / 2) + \delta_{s2} \sin^{4s}(\theta/2) }{\sin^4(\theta/2)} , \label{eq:lowfreq_grav_scat} 
\end{align}
where $\delta_{s2} = 1$ in the gravitational-wave case ($s=2$) and zero otherwise. Partial polarization is generated at order $O(a \omega)$ by the spin of the BH \cite{Dolan:2008kf}. The Rutherford-type divergence in the forward direction, of $d\sigma / d\Omega \sim 16 M^2 / \theta^4$, persists at high frequencies, due to the long-range $1/r$ potential of the Newtonian field. Of greater physical interest is the scattering through large angles, $\theta \gtrsim \pi / 2$, which leads to interference effects (orbiting and glories \cite{Matzner:1985rjn, Glampedakis:2001cx}) which are a diagnostic of the strong-field region of spacetime that harbors photon orbits, an ergoregion, and the event horizon.

The remainder of this paper is organized as follows. In Sec.~\ref{sec:linear_perturbations}, we briefly review the key equations for linear perturbations in Kerr spacetime. In Sec.~\ref{sec:scattering_cs}, we provide expressions for the differential scattering cross sections. In Secs.~\ref{sec:numerical_method}~and~\ref{sec:series_reduction}, we outline the numerical methods and the series reduction method, respectively, used to obtain the results presented in Sec.~\ref{sec:results}. We conclude with final remarks in Sec.~\ref{sec:remarks}. Throughout this paper we use natural units~($G=c=1$).

\section{Massless waves on Kerr background}\label{sec:linear_perturbations}
In the Boyer-Lindquist coordinates \{$t,r,\theta,\varphi$\} the Kerr metric reads 
\bea
{\d} s^2&=& -\f{\Delta}{\Sigma}\l({\d} t-a\sin^2\theta{\d} \varphi \r)^2 + \frac{\Sigma}{\Delta}{\d} r^2 + \Sigma\,{\d}\theta^2 + 
\nonumber\\
&& + \f{\sin^2\theta}{\Sigma}\l[(r^2+a^2){\d}\varphi-a{\d}t\r]^2,\label{eq:linelement}
\eea
with~$\Sigma\equiv r^2+a^2\cos^2\theta$ and $\Delta\equiv r^2-2Mr+a^2$. We restrict our attention to the case~$a^2 < M^2$, which corresponds to a rotating BH with two 
distinct horizons: an internal~(Cauchy) horizon located at~$r_{-}=M-\sqrt{M^2-a^2}$, and 
an external~(event) horizon at~$r_{+}=M+\sqrt{M^2-a^2}$.

On the Kerr background, massless waves are described by the Teukolsky master equation~\cite{teukolsky1972rotating}, which, when the field $\Upsilon_\mathfrak{s}$ is not sourced by any energy distribution, reads
\bea
\left[\frac{(r^{2}+a^{2})^{2}}{\Delta}-a^{2}\sin^{2}\theta\right]\frac{\partial^{2}\Upsilon_\mathfrak{s}}{\partial t^{2}}+\frac{4Mar}{\Delta}\frac{\partial^{2}\Upsilon_\mathfrak{s}}{\partial t\partial\varphi}\non+\left[\frac{a^{2}}{\Delta}-\frac{1}{\sin^{2}\theta}\right]\frac{\partial^{2}\Upsilon_\mathfrak{s}}{\partial\varphi^{2}}
-\Delta^{-\mathfrak{s}}\frac{\partial}{\partial r}\left(\Delta^{\mathfrak{s}+1}\frac{\partial\Upsilon_\mathfrak{s}}{\partial r}\right)\non-\frac{1}{\sin\theta}\frac{\partial}{\partial\theta}\left(\sin\theta\frac{\partial\Upsilon_\mathfrak{s}}{\partial\theta}\right)+(\mathfrak{s}^{2}\cot^{2}\theta-\mathfrak{s})\Upsilon_\mathfrak{s}\nonumber \\
-2\mathfrak{s}\left[\frac{a(r-M)}{\Delta}+\frac{i\cos\theta}{\sin^{2}\theta}\right]\frac{\partial\Upsilon_\mathfrak{s}}{\partial\varphi}\non-2\mathfrak{s}\left[\frac{M(r^{2}-a^{2})}{\Delta}-r-ia\cos\theta\right]\frac{\partial\Upsilon_\mathfrak{s}}{\partial t}=0,\label{eq:master_eq}
\eea
with $\mathfrak{s}$ being the spin weight of the field, and we have ${\mathfrak{s}}=0,\,\pm1/2,\,\pm1,\,\pm3/2,\,\pm2$, for scalar, spinorial, electromagnetic, Rarita-Schwinger~\cite{Gueven:1980be,Kamran:1985tp,TorresdelCastillo:1990aw}, and gravitational perturbations, respectively. 

In this paper we focus on bosonic waves; i.e.~we choose $\mathfrak{s}=0$, $-1$, and $-2$, noting the following: for scalar waves $\Upsilon_{0} \equiv \Phi$, where $\Phi$ is the scalar field; for electromagnetic waves $\Upsilon_{-1} \equiv \phi_2 (r-ia\cos\theta)^2$, where $\phi_2 \equiv F_{\mu \nu} \overline{m}^\mu n^\nu$ is a Maxwell scalar, $F_{\mu \nu}$ is the Faraday tensor, and $n^\mu$ and $\overline{m}^\mu$ are legs of Kinnersley's null tetrad \cite{teukolsky1972rotating}; for gravitational waves~(GWs) $\Upsilon_{-2} \equiv \psi_4 (r-ia\cos\theta)^4$, where $\psi_4=-C_{\alpha\beta\mu\nu}n^\alpha\overline{m}^\beta n^\mu\overline{m}^\nu$ is a Weyl scalar, and $C_{\alpha\beta\mu\nu}$ is the Weyl tensor, which in vacuum coincides with the Riemann tensor.

Using the standard ansatz~\cite{teukolsky1972rotating, teukolsky1973perturbations, press1973perturbations, teukolsky1974perturbations}
\be
\Upsilon_{\swlm}(t,\,r,\,\theta,\,\varphi)=\rtf_{\swlm}(r)\swf_{\swlm}(\theta)e^{-i(\omega t-m\varphi)},\label{eq:ansatz}
\ee
one can separate variables in the master equation [Eq.~\eqref{eq:master_eq}], obtaining the following pair of differential equations, 
\begin{align}
\l[\l(\Delta\f{{\d}^2}{{\d}r^2}+ (\sw+1)\Delta' \frac{{\d}}{{\d}r}\r)+\mathsf{V}_{\swlm}(r)\r]\rtf_{\swlm}=0,
\label{eq:radial_eq}
\\\l(\frac{{\d}^2}{{\d}\theta^2}+\cot\theta\frac{{\d}}{{\d}\theta}+\mathsf{A}_{\swlm}(\theta)\r)\swf_{\swlm}=0,\label{eq:angular_eq}
\end{align}
where
\begin{align}
\mathsf{V}_{\swlm}(r) \equiv \f{1}{\Delta}\left[K^{2}- i\mathfrak{s} \Delta' K\right]-\lambda_{\swlm}+4i\mathfrak{s}\omega r, \\
\mathsf{A}_{\swlm}(\theta)\equiv
2a\omega\l(m-\mathfrak{s}\cos\theta\r)-\f{\l(m+\mathfrak{s}\cos\theta\r)^2}{\sin^{2}\theta}\non+\lambda_{\swlm}+\mathfrak{s} - a^2 \omega^2 \sin^2 \theta,
\end{align}
with $K\equiv(r^{2}+a^{2})\omega-am$. The angular functions, $\swf$, satisfying Eq.~\eqref{eq:angular_eq} are known in the literature as the spin-weighted spheroidal functions (or harmonics), and the quantities $\lambda_{\swlm}$ are their eigenvalues. Hereafter, we refer to Eq.~\eqref{eq:radial_eq} as the radial Teukolsky equation (RTE).

The boundary conditions for the RTE are
\be
\rtf_{\swlm}\sim\begin{cases}
	\mathbb{T}_{\swlm}e^{-\imath\omtil x}\Delta^{-\sw} , & x\rightarrow -\infty,\\
	\mathbb{I}_{\swlm}\f{e^{-\imath\omega x}}{r}+\mathbb{R}_{\swlm}\f{e^{\imath\omega x}}{r^{(2\sw+1)}}, & r\rightarrow 
	+\infty,\label{eq:ingoing_sol}
\end{cases} 
\ee
where $\omtil\equiv\omega-am/2Mr_+$, and $x$ is the tortoise coordinate in 
Boyer-Lindquist coordinates, defined by~\cite{Ottewill:2000qh} 
\begin{align}
x &\equiv r+\frac{1}{r_+ - r_-}\l[ (r_+^2+a^2)\ln|r-r_+| \r. \nn \\ & \quad \quad \quad \quad \quad \quad \quad  \l.  -(r_-^2+a^2)\ln|r-r_-|\r]. 
\end{align}

\section{Scattering cross section}\label{sec:scattering_cs}
The differential scattering cross section can be expressed as follows:
\be
\frac{d\sigma}{d\Omega}=|\mathfrak{f}(\theta)|^2+|\mathfrak{g}(\theta)|^2.
\label{eq:cross_section}
\ee
Using the partial-wave method, the helicity-conserving $\mathfrak{f}(\theta)$~\cite{Glampedakis:2001cx,Dolan:2008kf} and helicity-reversing $\mathfrak{g}(\theta)$~\cite{Dolan:2008kf} scattering amplitudes can be written as
\begin{widetext}
\begin{align}
\mathfrak{f}(\theta)=\begin{cases}
\frac{2\pi}{\imath\omega}\sum_{l=0}^{^{\infty}}\swf_{0 \omega l0}(0)\swf_{0\omega l0}(\theta)\left(e^{2\imath\delta_{0\omega l0}^{(s=0)}}-1\right), & \text{for}\qquad s=0,\\
\frac{2\pi}{\imath\omega}\sum_{l=1}^{^{\infty}}\swf_{-1\omega l1}(0)\swf_{-1\omega l1}(\theta)\left(e^{2\imath\delta_{-1\omega l1}^{(s=1)}}-1\right), & \text{for}\qquad s=1,\\
\frac{\pi}{\imath\omega}\sum_{P=\pm1}\sum_{l=2}^{^{\infty}}\swf_{-2\omega l2}(0)\swf_{-2\omega l2}(\theta)\left(e^{2\imath\delta_{-2 \omega l2P}^{(s=2)}}-1\right), & \text{for}\qquad s=2,
\label{eq:hel_con}
\end{cases}
\end{align}
and
\begin{align}
\mathfrak{g}(\theta)=\begin{cases}
\frac{\pi}{\imath\omega}\sum_{P=\pm1}\sum_{l=2}^{^{\infty}}P\,(-1)^{l}\, \swf_{-2\omega l2}(0)\swf_{-2\omega l2}(\pi-\theta) \left(e^{2\imath\delta_{-2\omega l2P}^{(s=2)}}-1\right), & \text{for}\qquad s=2,\\
0, & \text{otherwise},
\label{eq:hel_rev}
\end{cases}
\end{align}
\end{widetext}
where $e^{2\imath\delta_{0\omega l0}^{(s=0)}}$, $e^{2\imath\delta_{-1\omega l1}^{(s=1)}}$, and $e^{2\imath\delta_{-2 \omega l2 P}^{(s=2)}}$ are the phase shifts which can be computed by integrating the RTE, and $\theta$ denotes the scattering angle. We point out that for GWs $(s=2)$ there is a sum over parities $(P=\pm1)$.

The phase shifts are given explicitly by
\begin{subequations}
\beq
&e^{2\imath\delta_{0\omega l0}^{(s=0)}}=(-1)^{l+1}\frac{\mathbb{R}_{0\omega l0}}{\mathbb{I}_{0\omega l0}},
\label{eq:scalar_phaseshift}\\
&e^{2\imath\delta_{-1\omega l1}^{(s=1)}}=(-1)^{l+1}\frac{\mathcal{B}_{\omega l1}}{4\omega^2}\frac{\mathbb{R}_{-1\omega l1}}{\mathbb{I}_{-1 \omega l1}},
\label{eq:EM_phaseshift}\\
&e^{2\imath\delta_{-2 \omega l2P}^{(s=2)}}=(-1)^{l+1}\l(\frac{\text{Re}(\mathcal{C})+12i\omega MP}{16\omega^4}\r)\frac{\mathbb{R}_{-2\omega l2}}{\mathbb{I}_{-2\omega l2}},
\label{eq:GWs_phaseshift}
\eeq
\end{subequations}
where $\mathcal{B}_{\omega l1}^2=\lambda_{-1\omega l1}^2+4a\omega-4a^2\omega^2$ and $[\text{Re}(\mathcal{C})]^2=[(\lambda_{-2\omega l2}+2)^2+4a\omega-4a^2\omega^2](\lambda_{-2\omega l2}^2+36a\omega-36a^2\omega^2)+(2\lambda_{-2\omega l2}+3)(96a^2\omega^2-48a\omega)-144a^2\omega^2$. Due to the parity dependence in Eq.~\eqref{eq:GWs_phaseshift}, the helicity-reversing amplitude $\mathfrak{g}(\theta)$ is nonzero for GWs~\cite{Dolan:2008kf}.

\section{Numerical method}\label{sec:numerical_method}
From Eqs.~\eqref{eq:hel_con} and~\eqref{eq:hel_rev} we note that we need to obtain the spin-weighted spheroidal harmonics (and their eigenvalues) and the phase shifts [Eqs.~\eqref{eq:scalar_phaseshift}--\eqref{eq:GWs_phaseshift}] in order to use the formula given by Eq.~\eqref{eq:cross_section}. An additional problem impeding the calculation of the scattering cross section is the lack of convergence of the partial-wave series given in Eq.~\eqref{eq:hel_con}.

We obtain the spin-weighted spheroidal harmonics and their eigenvalues via spectral decomposition using the description outlined in Refs.~\cite{Dolan:2008kf,Leite:2017zyb,Hughes:1999bq,Cook:2014cta}, in which $\swf_{\swlm}$ is written as a sum of spin-weighted spherical harmonics,
\be
\swf_{\sw \omega jm} = \sum^{\infty}_{l=l_{\text{min}}} b_{\omega jl|\sw|}Y_{\mathfrak{s} lm}  \label{eq:spec}
\ee
where $l_{\text{min}} = \text{max}(|\mathfrak{s}|, |m|)$. In order to obtain the phase shifts, we  numerically integrate the RTE using the numerical schemes detailed in Ref.~\cite{Leite:2017zyb}. 

The long-ranged characteristic of the gravitational interaction leads to a divergence in the amplitude $\mathfrak{f}(\theta)$ at $\theta=0$~\cite{Dolan:2008kf,Glampedakis:2001cx}. This physical divergence leads to a lack of convergence in the series representation in Eq.~\eqref{eq:hel_con} for \emph{any} value of $\theta$. In order to improve the convergence properties of the series, we develop and apply a series reduction method, as described in Sec.~\ref{sec:series_reduction}. 

In a numerical calculation of $d\sigma/d\Omega$, there are several sources of numerical error, including (i) global truncation error in the numerical integration of the RTE; (ii) fitting error in matching the radial solutions to truncated series in the asymptotic regimes ($x \rightarrow \pm \infty$) to obtain phase shifts; (iii) numerical error in application of the series reduction method; (iv) truncation error in terminating an infinite sum at an appropriate $l_{\text{max}}$. We find that (i) and (ii) are not significant, whereas (iii) and (iv) put a practical limit on the numerical accuracy achieved. For comparison purposes, sample data are given in Table \ref{tbl:data}, along with an error estimate found from the results of applying two and three iterations of the series reduction method (below) for $l_{\text{max}} = 60$. If required, more precise results could be obtained by 
increasing $l_{\text{max}}$.

\section{Series reduction}\label{sec:series_reduction}
The scheme described in this section is inspired by a method developed for numerical computations of Coulombian scattering~\cite{Yennie_1954:prd95_500}, which has been successfully employed to compute BHs scattering cross sections~\cite{Dolan:2008kf}.

First, let us note that one can rewrite the partial-wave series $\mathfrak{f}(\theta)$ [Eq.~\eqref{eq:hel_con}] and $\mathfrak{g}(\theta)$ [Eq.~\eqref{eq:hel_rev}] in the following generic form:
\be
F(\theta)=\sum_{l=|\sw|}^{+\infty}F_{l\omega} \,\swf_{\sw\omega l |\sw|}(\theta).\label{eq:partialwave_generic_series}
\ee 
Using the spin-weighted spheroidal harmonics spectral decomposition, one can show that the sum over spin-weighted spheroidal harmonics [Eq.~\eqref{eq:partialwave_generic_series}] can be rewritten as
\be
F(\theta)=\sum_{j=|\sw|}^{+\infty}\mathtt{F}_j \,Y_{\sw j |\sw|}(\theta),
\ee 
where we have defined $\mathtt{F}_j\equiv\sum_{l=|\sw|}^{+\infty}F_{l\omega} \,b_{\omega jl|\sw|}$, with $b_{\omega jl|\sw|}$ being the spectral decomposition coefficients of Eq.~\eqref{eq:spec}.

The series reduction technique involves defining a new series
\be
F(\theta)=(1-\cos\theta)^{-n}\sum_{j=|\sw|}^{+\infty}\mathtt{F}{_j^{(n)}} \,Y_{\sw j |\sw|}(\theta),
\label{eq:series_reduction}
\ee 
which has more amenable convergence properties. Noting that $\mathtt{F}_j=\mathtt{F}{_j^{(0)}}$, the coefficients $\mathtt{F}{_j^{(n)}}$ can be obtained from the following recurrence relations:
\be
\mathtt{F}{_j^{(n+1)}}=\l(1-\mathcal{X}_{\sw j |\sw|}\r)\mathtt{F}{_j^{(n)}}-\l[\mathcal{Y}_{\sw (j-1) |\sw|}\mathtt{F}{_{j-1}^{(n)}}+\mathcal{Z}_{\sw (j+1) |\sw|}\mathtt{F}{_{j+1}^{(n)}}\r],
\label{eq:series_reduction_coeff}
\ee
where
\begin{eqnarray}
\mathcal{Y}_{{\sw}jm} & = & 
\sqrt{\frac{\left(j+1\right)^{2}-m^{2}}{
		\left(2j+1\right)\left(2j+3\right)}}\sqrt{\frac{\left(j+1\right)^{2}
		-{\sw}^{2}}{\left(j+1\right)^{2}}},\\
\mathcal{Z}_{{\sw}jm} & = & \begin{cases}
\sqrt{\frac{j^{2}-{\sw}^{2}}{j^{2}}}\sqrt{\frac{j^{2}-m^{2}}{4j^{2}-1}}, & 
\text{for}\: j\neq0,\\
0, & \text{for}\: j=0,
\end{cases}\\
\mathcal{X}_{{\sw}jm} & = & \begin{cases}
-\frac{m{\sw}}{j\left(j+1\right)}, & 
\text{for}\: j\neq0\:\text{and}\: {\sw}\neq0,\\
0, & \text{for}\: j=0\:\text{or}\: {\sw}=0.
\end{cases}
\end{eqnarray}
The recurrence relations given by Eq.~\eqref{eq:series_reduction_coeff} are obtained using the properties of the spin-weighted spherical harmonics. We compute the scattering amplitudes $\mathfrak{f}(\theta)$ and $\mathfrak{g}(\theta)$ with the help of Eq.~\eqref{eq:series_reduction}, with the typical choice of $n=2$. Moreover, we truncate our series at finite values $l_{\text{max}}$ and $j_{\text{max}}$, depending on the value of the coupling $\omega M$.

\section{Numerical Results}\label{sec:results}

\begin{figure*}
	\begin{center}
		\includegraphics[width=0.5\columnwidth]{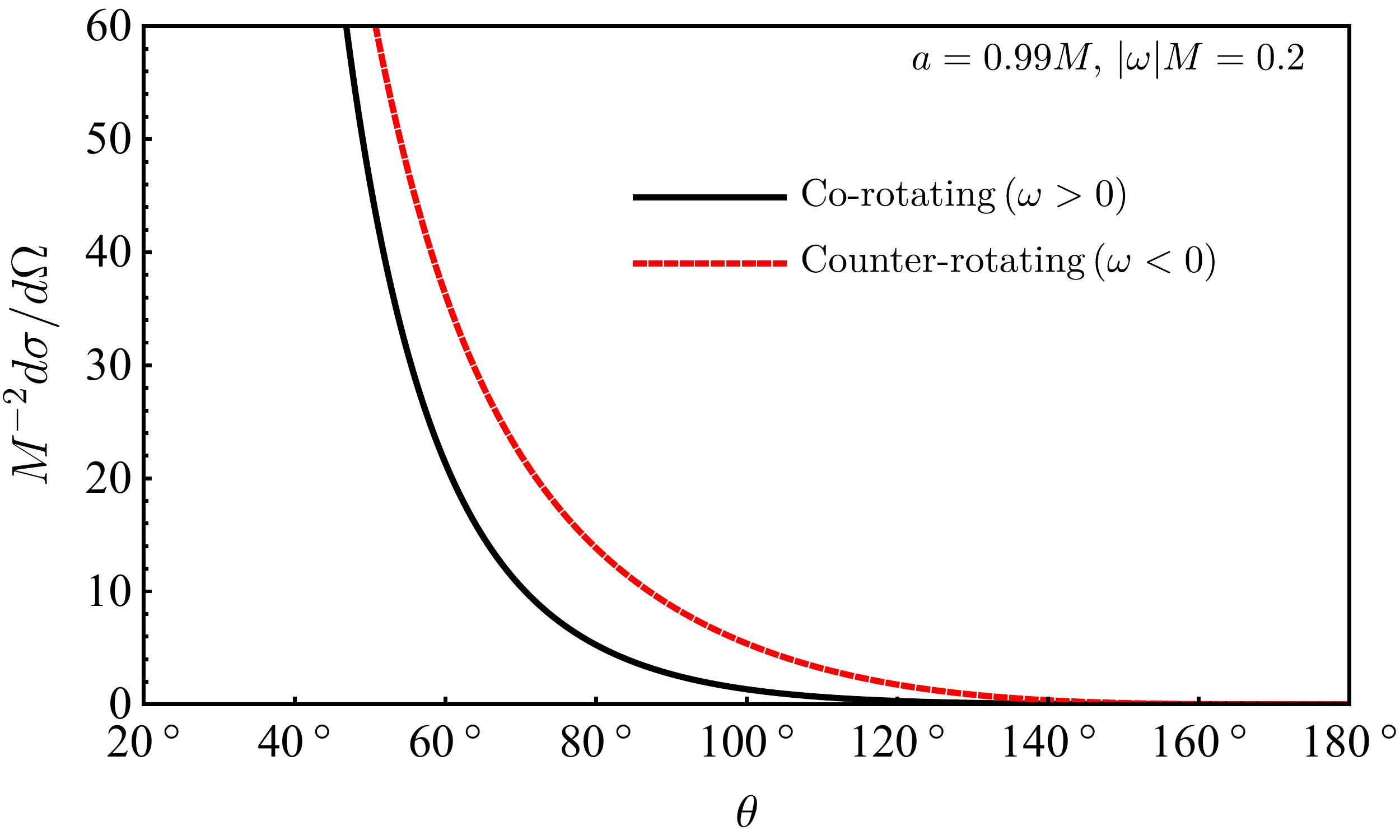}\includegraphics[width=0.5\columnwidth]{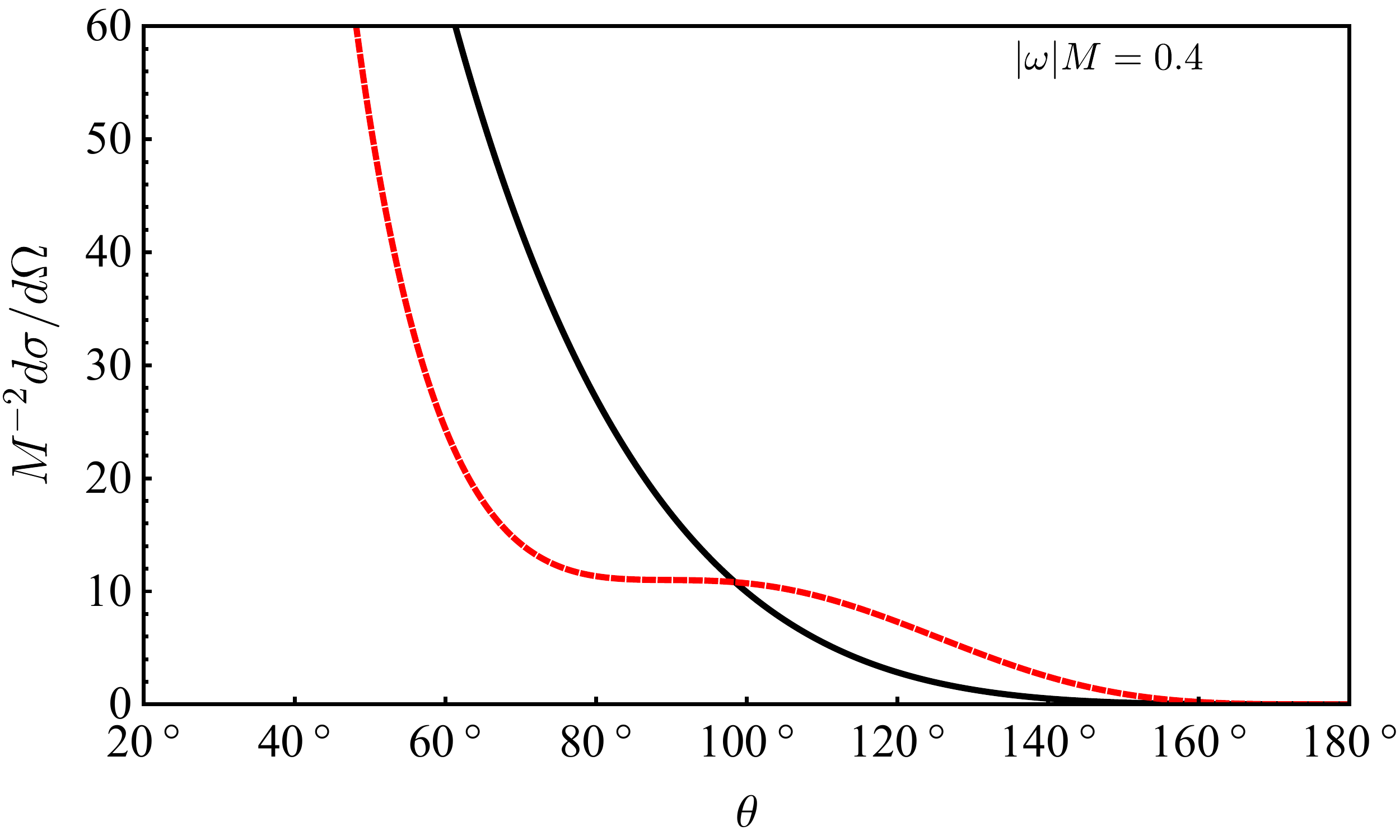}\\
		\includegraphics[width=0.5\columnwidth]{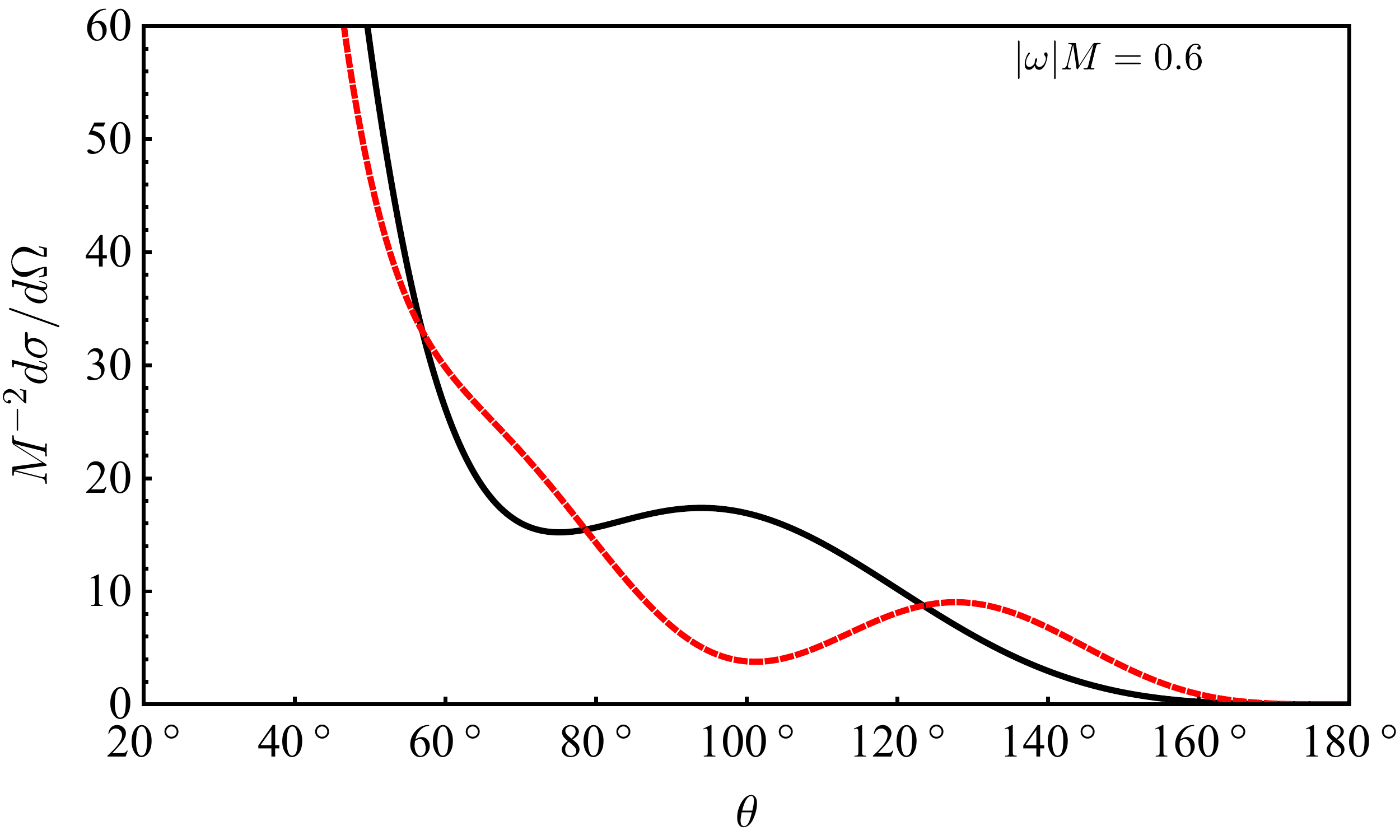}\includegraphics[width=0.5\columnwidth]{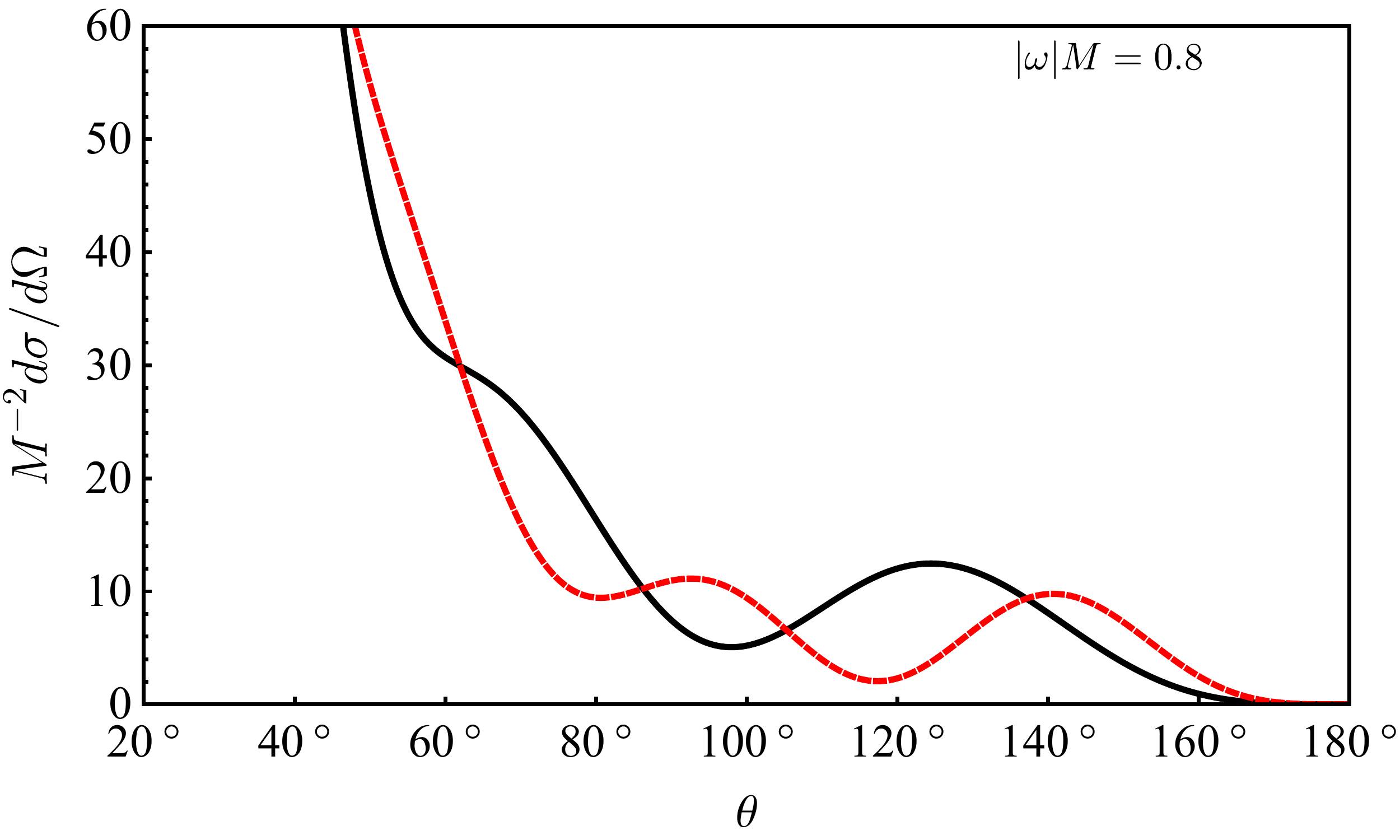}
		\caption{Low-frequency electromagnetic Kerr scattering cross sections for on-axis incidence. We consider corotating~($\omega M>0$) and counter-rotating~($\omega M<0$) polarized electromagnetic waves impinging upon a rapidly rotating ($a=0.99M$) BH.}
		\label{fig:EM_low}
	\end{center}
\end{figure*}

Figure \ref{fig:EM_low} shows the scattering cross section for low-frequency ($0.2 \le |\omega| M \le 0.8$) electromagnetic waves ($s=1$) impinging upon a rapidly rotating Kerr BH~($a=0.99M$). Corotating~($a \omega >0$) and counter-rotating~($a \omega <0$) waves are scattered in a different way, due to the coupling between the helicity of the field and the rotation of the BH. Thus, a partial polarization is generated in an unpolarized beam by the frame-dragging of spacetime. In the backward direction~($\theta=180^{\circ}$), the cross section is zero for both co- and counter-rotating polarizations.

\begin{figure*}
	\begin{center}
		\includegraphics[width=0.5\columnwidth]{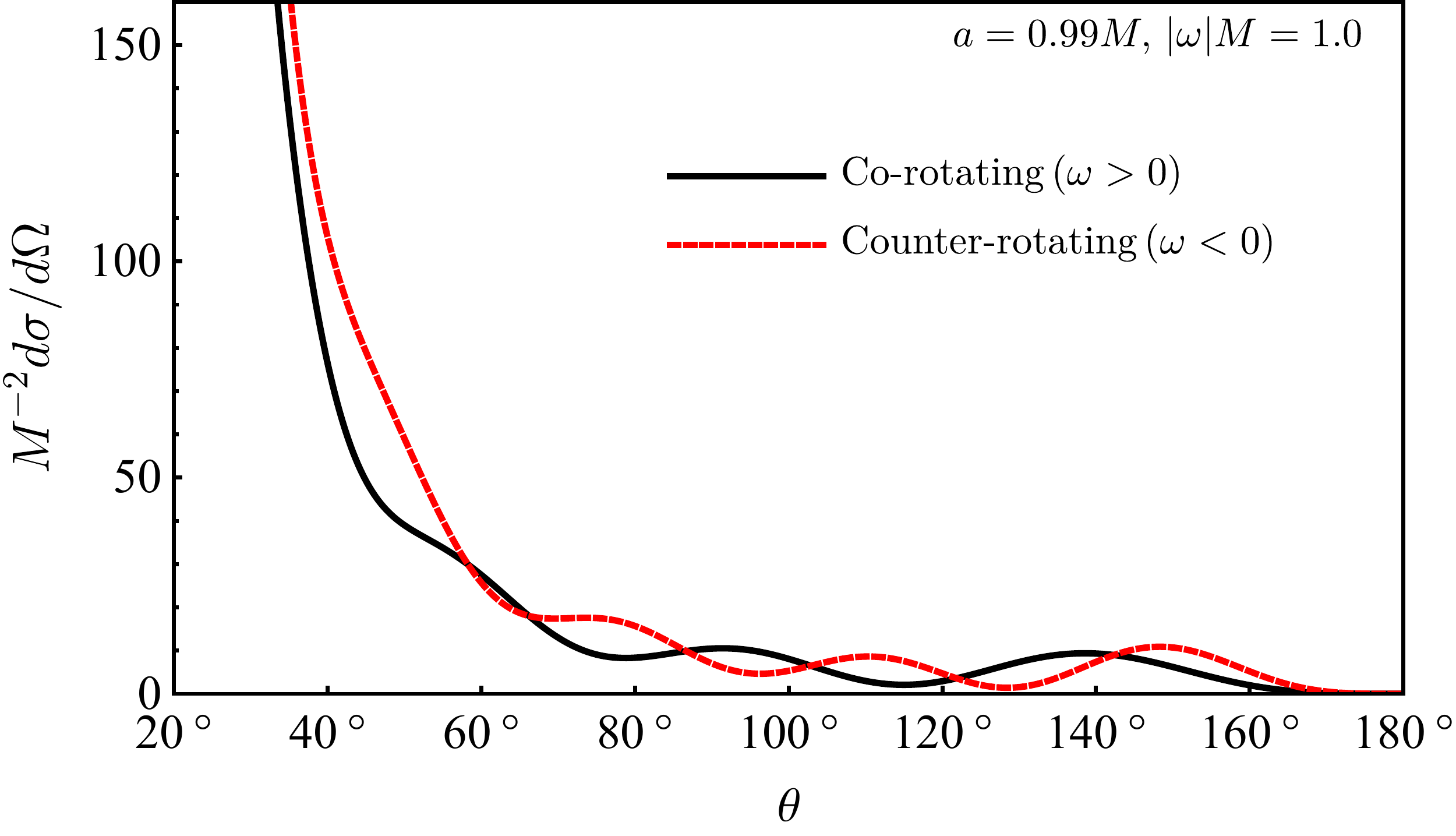}\includegraphics[width=0.5\columnwidth]{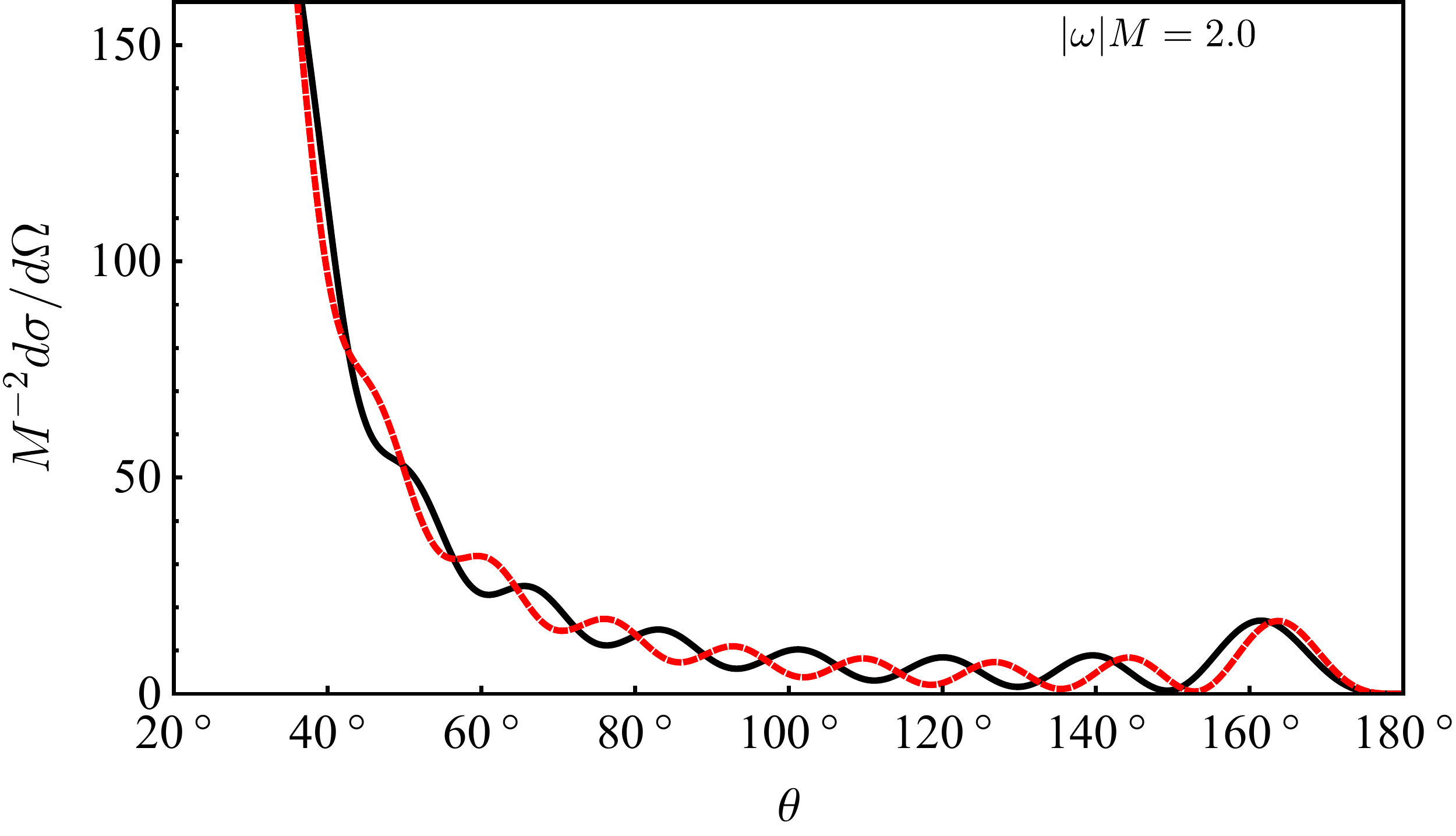}\\
		\includegraphics[width=0.5\columnwidth]{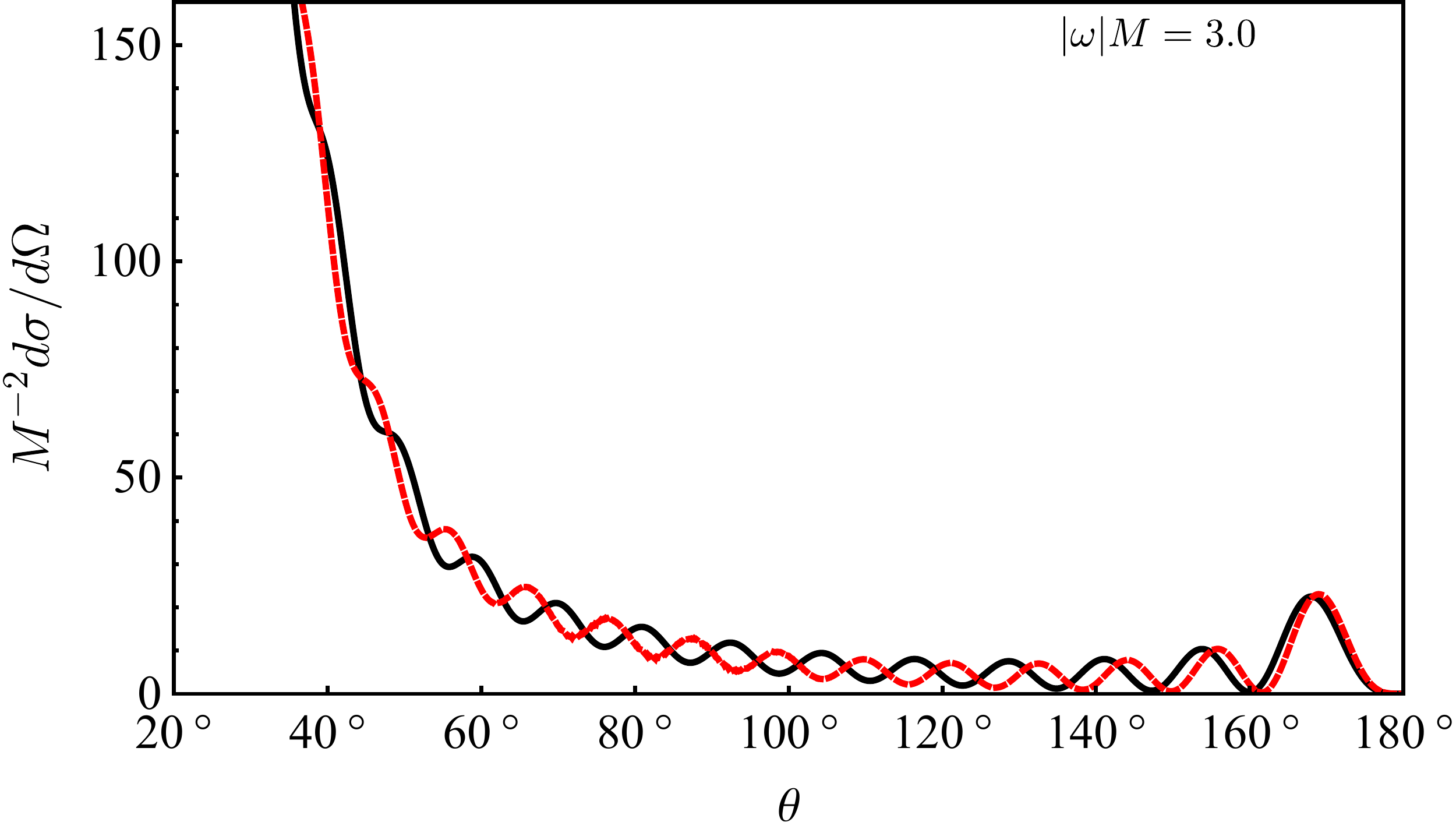}\includegraphics[width=0.5\columnwidth]{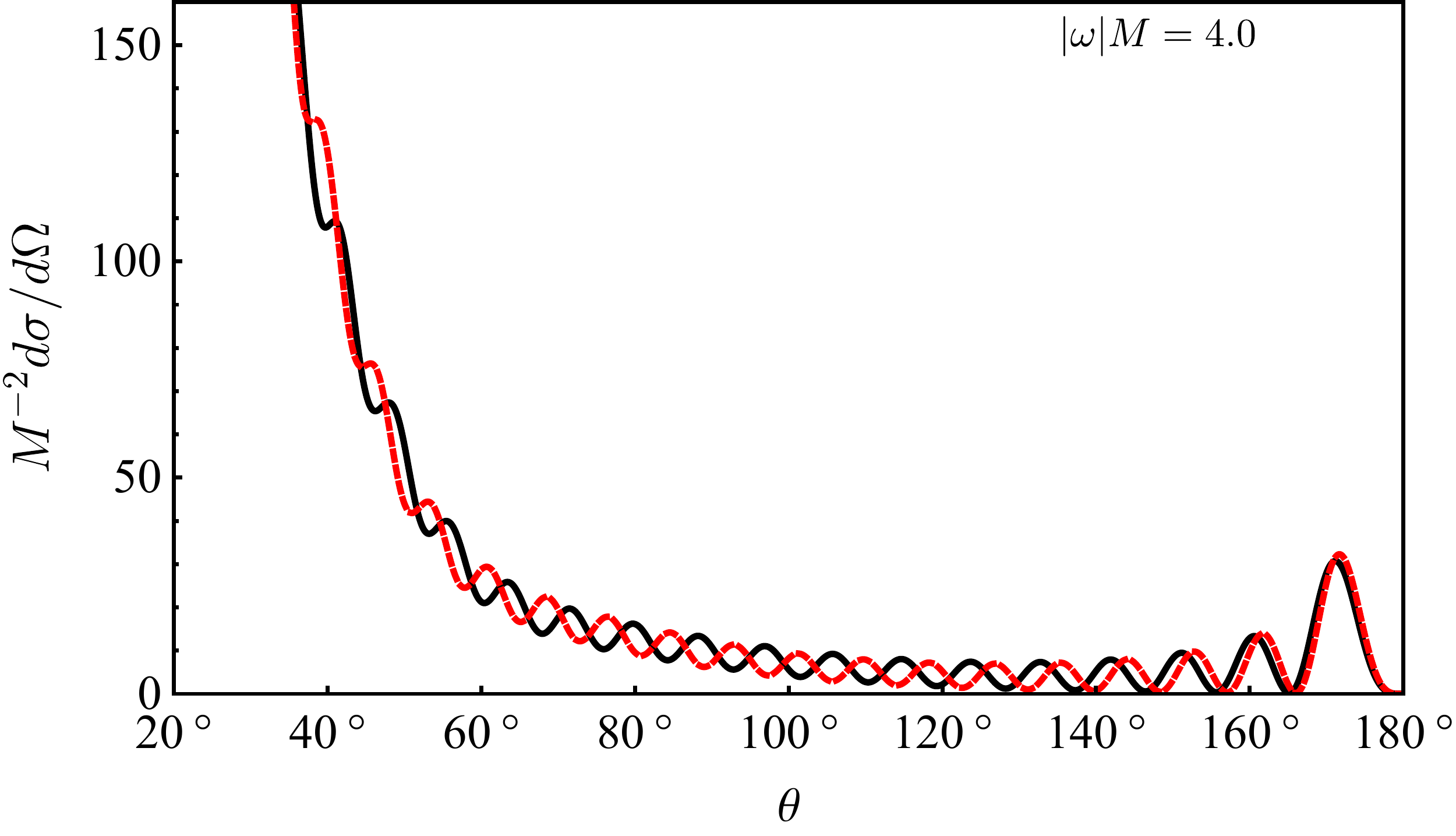}
		\caption{High-frequency Kerr scattering cross sections for on-axis incidence, showing corotating~($a \omega >0$) and counter-rotating~($a \omega <0$) polarized electromagnetic waves impinging upon a rapidly rotating ($a=0.99M$) BH.}
		\label{fig:EM_high}
	\end{center}
\end{figure*}

Figure \ref{fig:EM_high} exhibits the scattering cross section for high-frequency electromagnetic waves, demonstrating the existence of orbiting oscillations. The semiclassical interpretation is that these oscillations arise due to constructive or destructive interference between a pair of rays which scatter through $\theta$ and $2 \pi - \theta$. The angular width of the orbiting oscillation diminishes in inverse proportion to the frequency, as expected. We note that the co- and counter-rotating oscillations have subtly different angular widths.

\begin{figure}
\begin{center}
	\includegraphics[width=0.5\columnwidth]{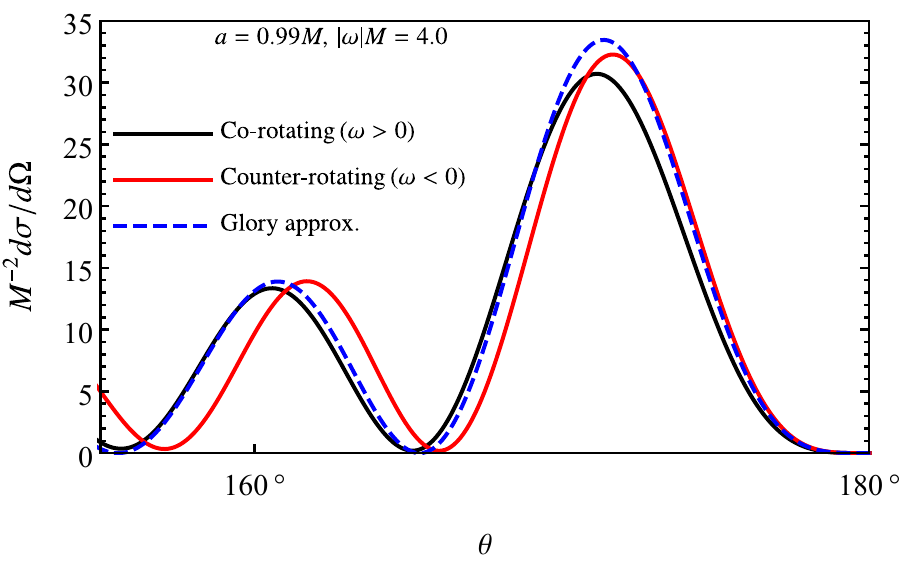}
	\caption{
	Electromagnetic scattering for $|\omega| M = 4$ and  $a = 0.99M$. The dotted blue line shows the analytical approximation to the glory, Eq.~(\ref{eq:glory}), with $s=1$, $b_g = 5.0925M$, and $d b_g / d\theta = -0.2209M$ \cite{Dolan:2008kf}. The lines show the cross sections of the corotating (black) and counter-rotating (red) polarizations obtained by our numerical method. 
	}
	\label{fig:glory}
\end{center}
\end{figure}

The cross section in the vicinity of the axis shows a bright spot ($s=0$) or ring ($s > 0$) known as a glory, which arises from  interference between a one-parameter family of rays that originate from an annulus centered on the axis. Matzner \emph{et al.} \cite{Matzner:1985rjn} derived the approximation
\be
\frac{d\sigma}{d\Omega} \approx 2 \pi M \omega b_g^2 \left| \frac{d b}{d \theta} \right|_{\theta=\pi} J_{2s}^2 (\omega b_g \sin \theta) , \label{eq:glory}
\ee
where $b_g$ is the impact parameter for the ray scattered through $180^\circ$ and $J_{2s}(\cdot)$ is a Bessel function.  Equation (\ref{eq:glory}) is valid for any spin $s$, but it does not distinguish between the co- and counter-rotating polarizations. In Fig.~\ref{fig:glory} we compare our numerical results with the approximation (\ref{eq:glory}). We see that the two polarizations reach maxima at slightly different scattering angles, with the counter-rotating polarization slightly closer to the axis than the corotating polarization. Although Eq.~(\ref{eq:glory}) is a robust approximation, it does not include this effect.

\begin{figure*}
 \begin{center}

	\includegraphics[width=0.5\columnwidth]{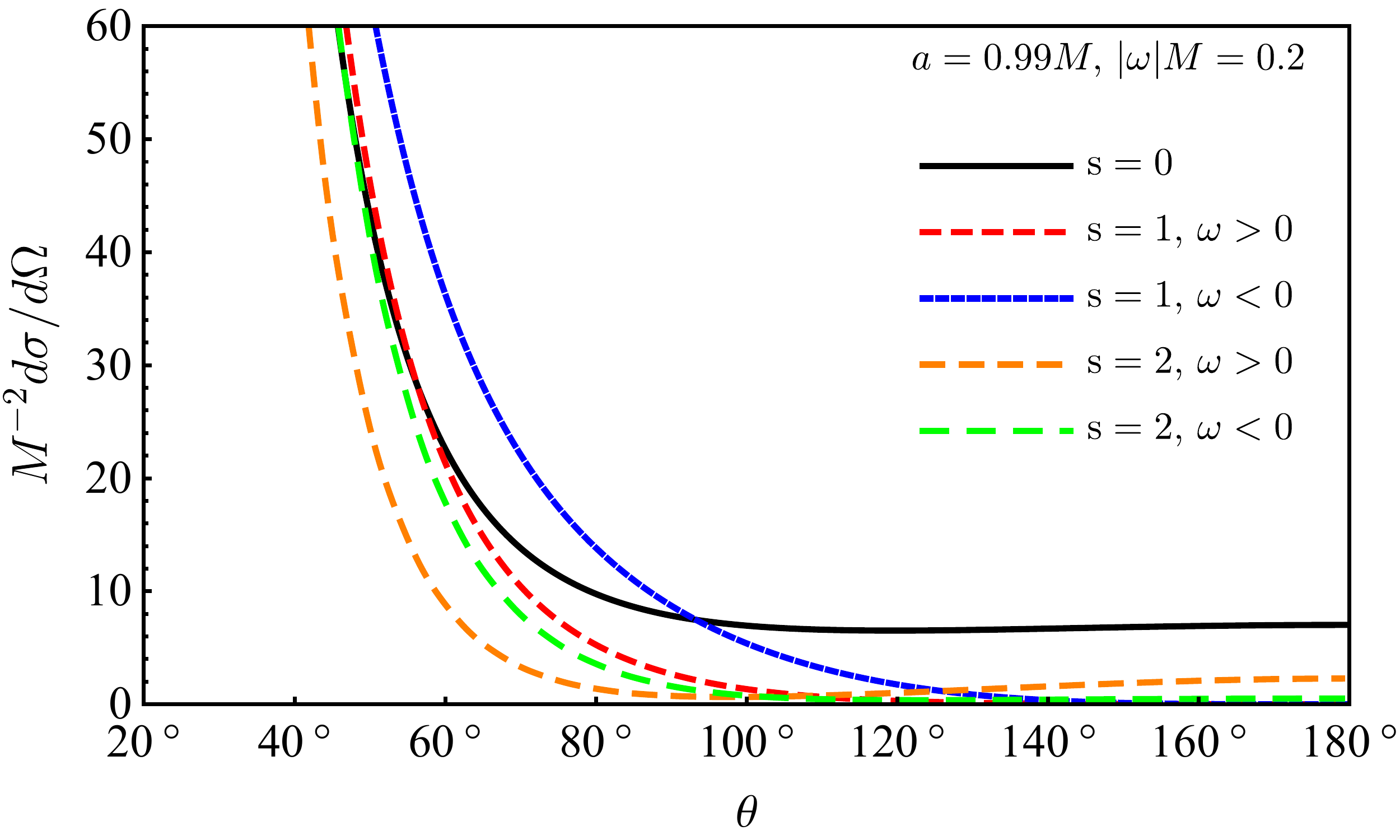}\includegraphics[width=0.5\columnwidth]{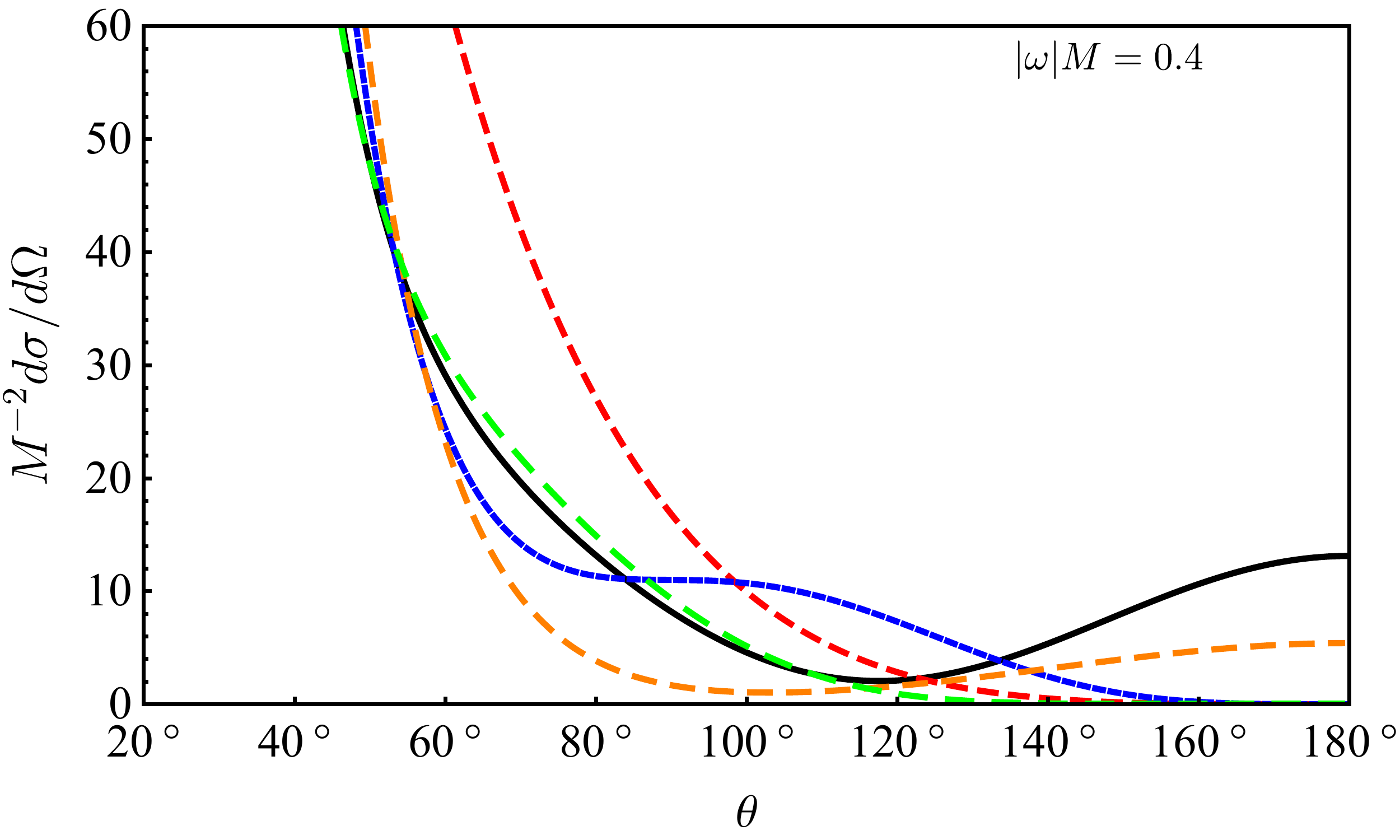}\\
	\includegraphics[width=0.5\columnwidth]{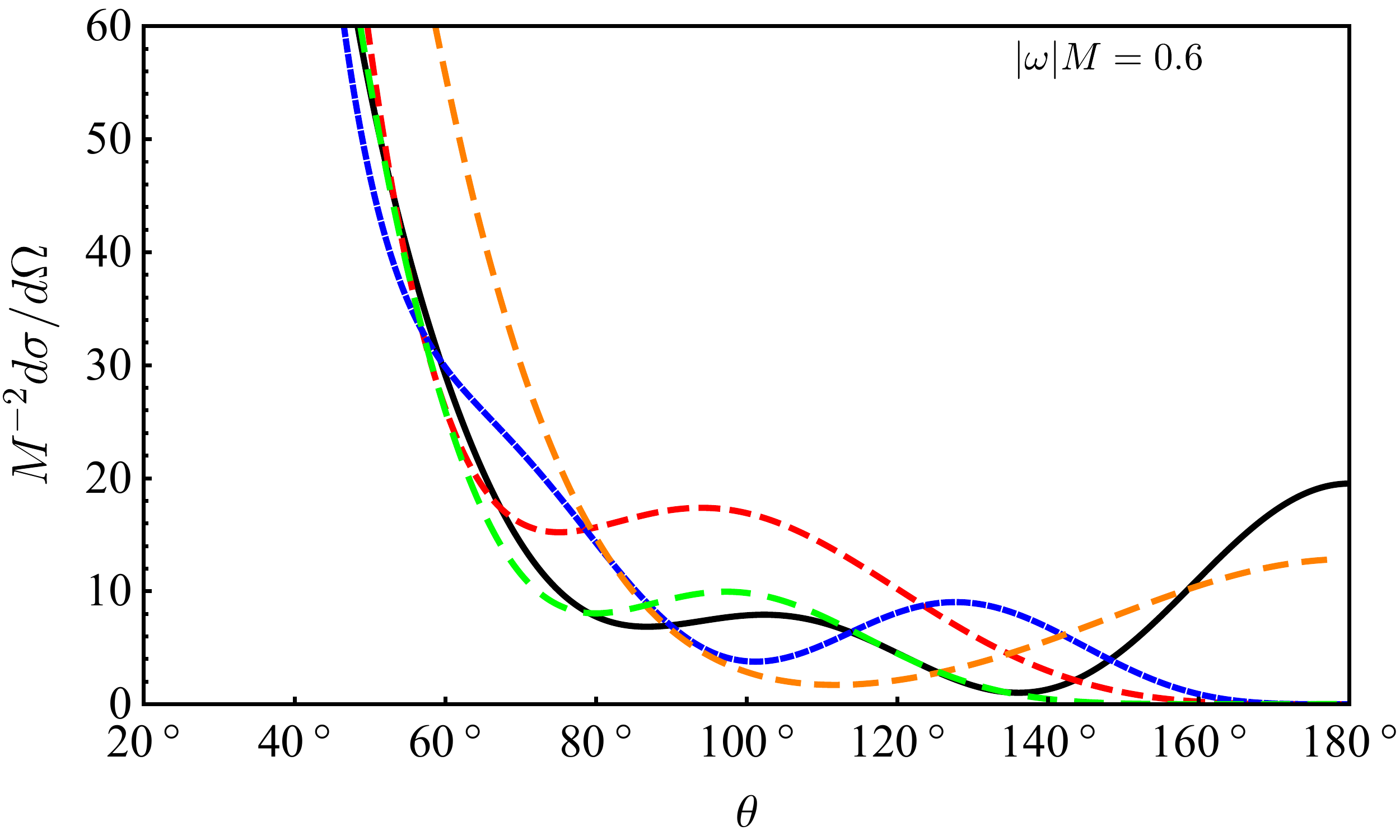}\includegraphics[width=0.5\columnwidth]{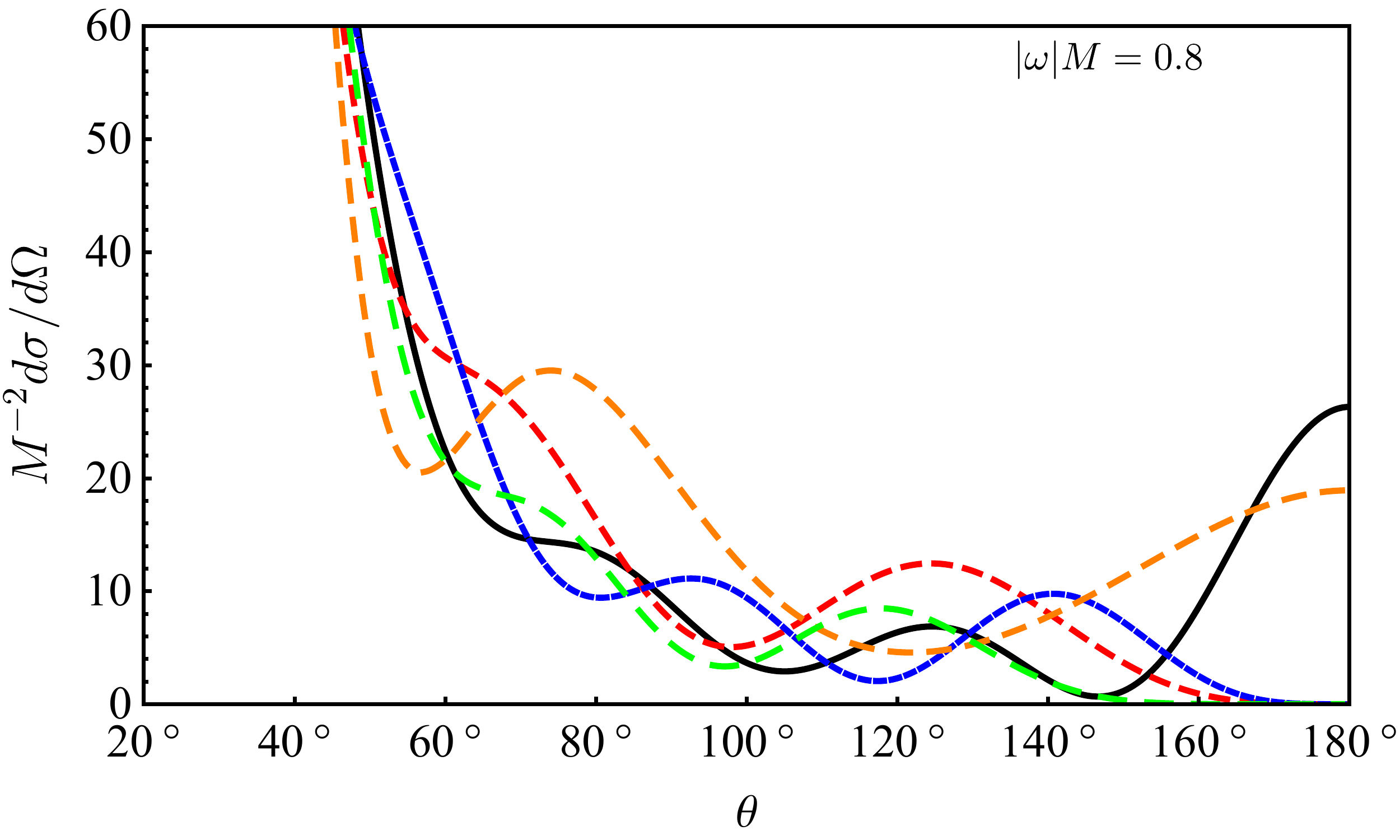}
	\caption{Long-wavelength ($\omega M < 1$) Kerr scattering cross sections for on-axis incidence, showing both circular polarizations that are co-rotating ($\omega > 0$) and counter-rotating ($\omega < 0$) with the spin of a 
	rapidly rotating ($a=0.99M$) BH.
	}
	\label{fig:all_spin_lowfreq}
	\end{center}
\end{figure*}

\begin{figure*}
\begin{center}
	\includegraphics[width=0.5\columnwidth]{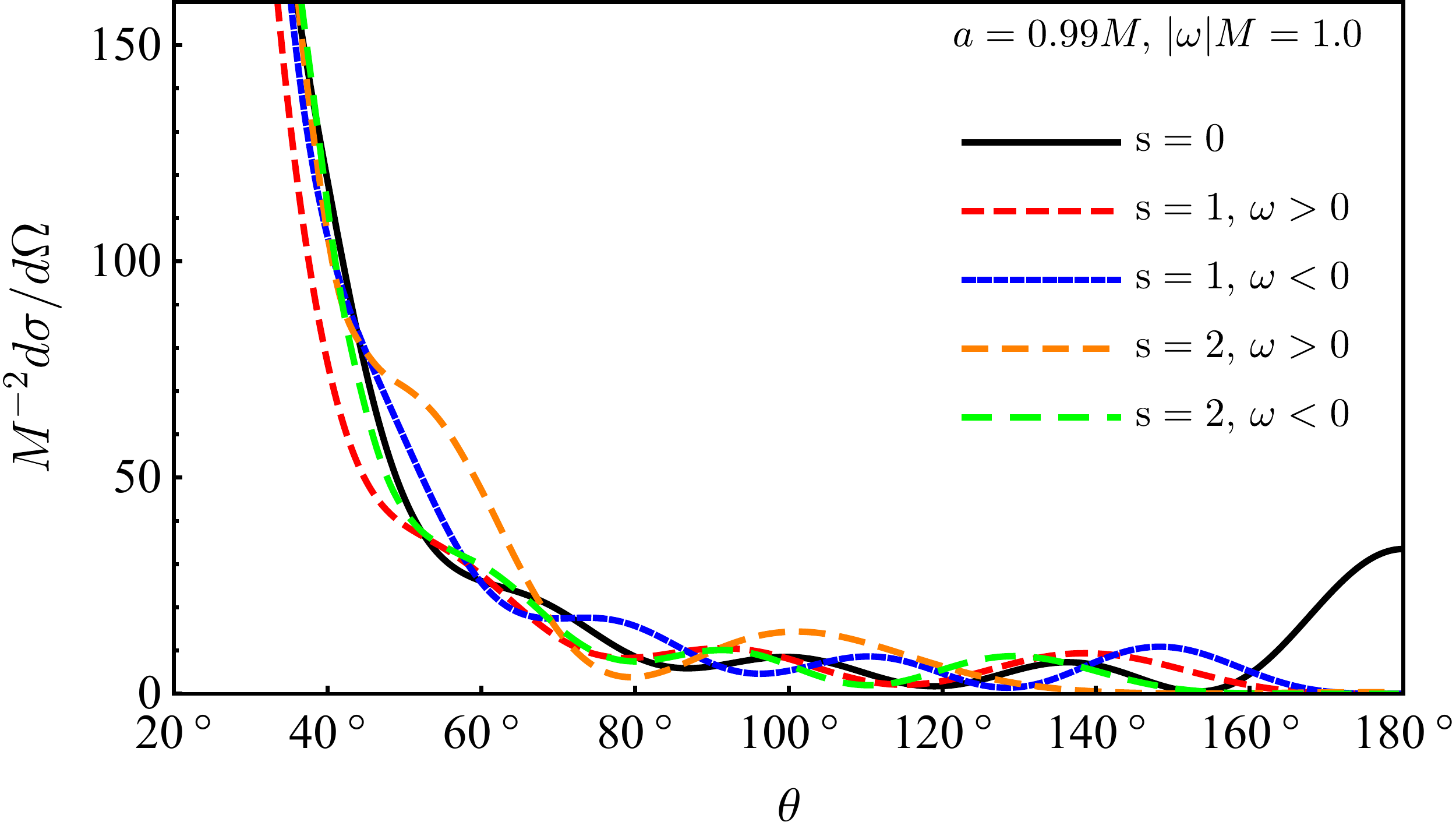}\includegraphics[width=0.5\columnwidth]{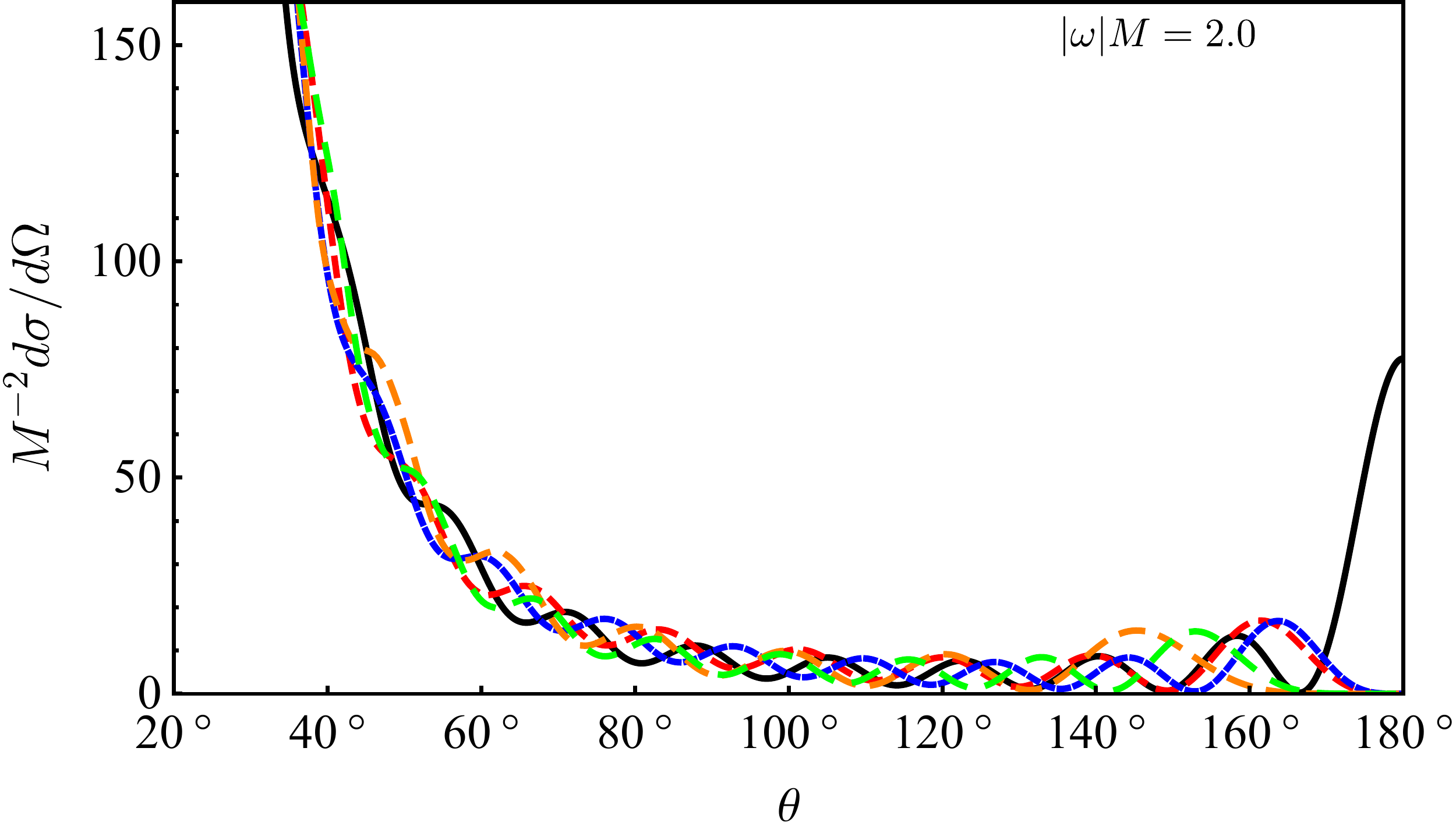}\\
	\includegraphics[width=0.5\columnwidth]{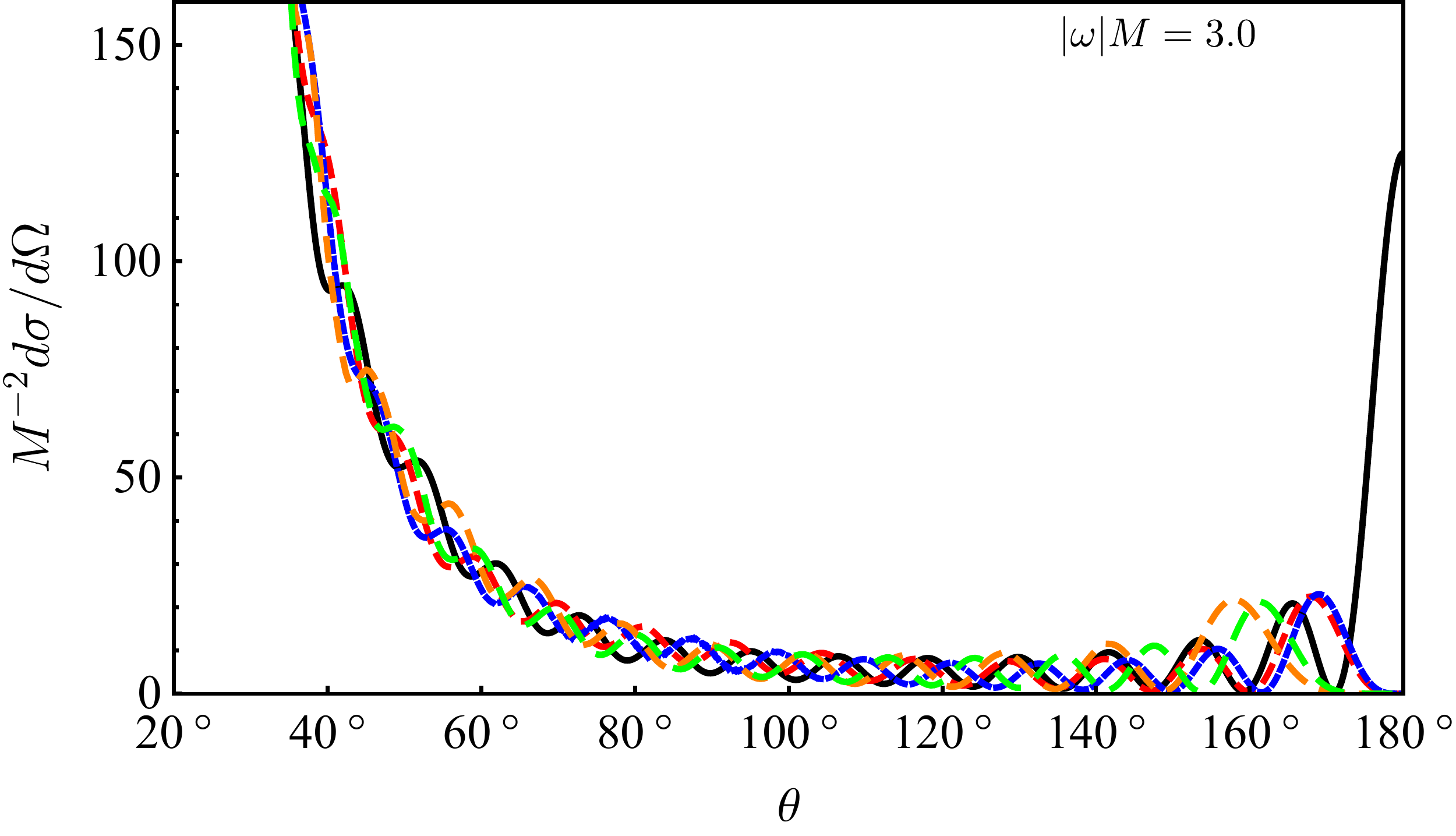}\includegraphics[width=0.5\columnwidth]{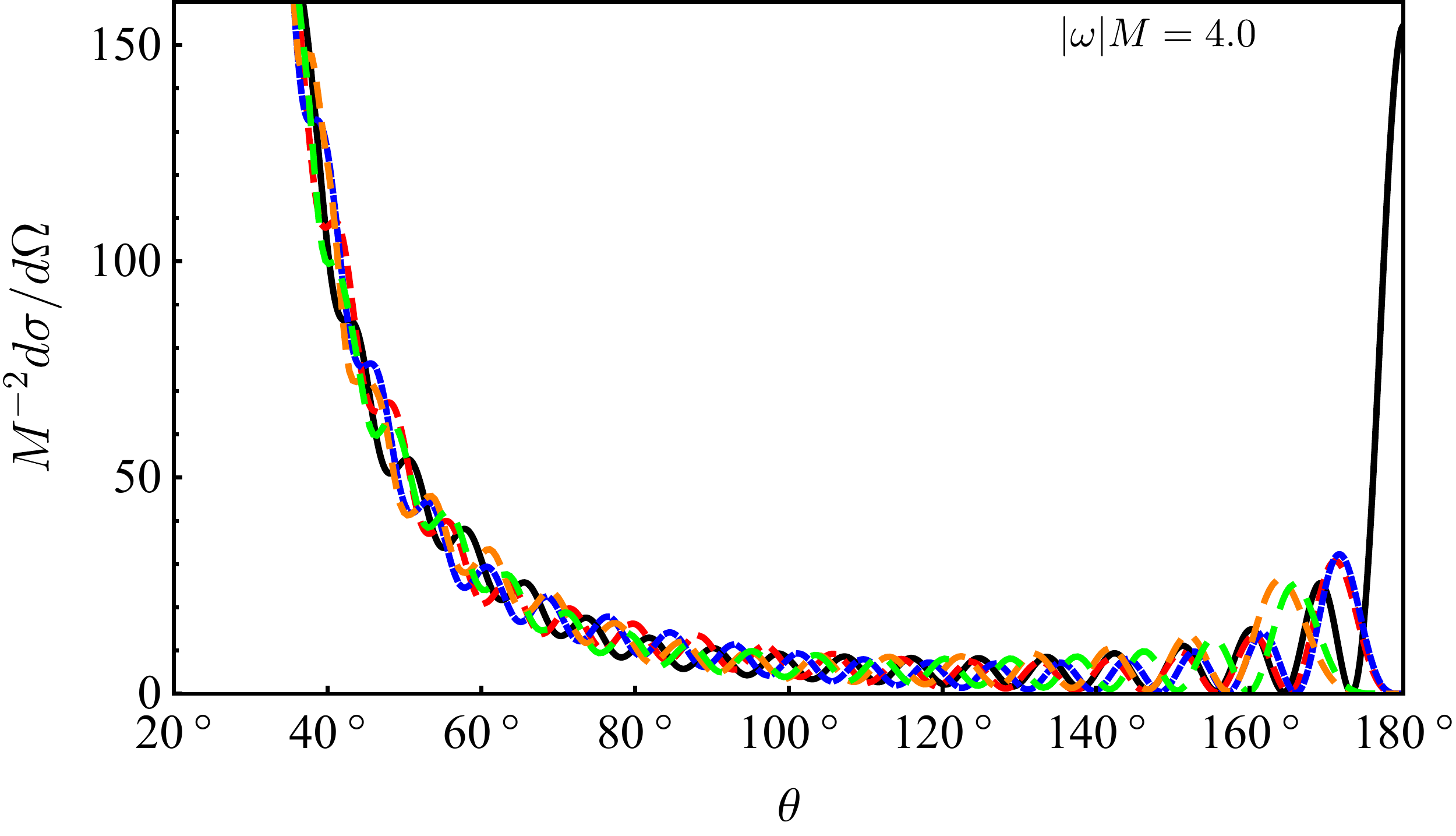}
	\caption{Short-wavelength ($\omega M\ge 1$) Kerr scattering cross sections for on-axis incidence.
	We consider scalar ($s=0$), electromagnetic ($s=1$), and gravitational ($s=2$) plane waves
	impinging on-axis upon a rapidly rotating ($a=0.99M$) BH.
	}
	\label{fig:all_spin_highfreq}
\end{center}
\end{figure*}

Figures \ref{fig:all_spin_lowfreq} and \ref{fig:all_spin_highfreq} show the scattering cross sections for massless fields of spin $s=0$, $1$, and $2$, corresponding to scalar ($s=0$), electromagnetic ($s=1$), and gravitational ($s=2$) waves impinging along the rotation axis of a rapidly rotating BH ($a=0.99M$). Figure \ref{fig:all_spin_lowfreq} examines the  long-wavelength regime ($0.2\leq|\omega|M\leq0.8$). For cases $s > 0$, the cross sections depend on the circular polarization of the wave, with the corotating ($M\omega >0$) and counter-rotating polarizations ($M \omega < 0$) scattering differently. For EM waves, the cross section is  zero in the backward direction ($\theta=180^{\circ}$) as the parallel transport of spin leads to destructive interference here. For scalar waves, there is a nonzero backward-scattered flux. For GWs, a nonzero backward flux arises from the helicity-reversing term $|\mathfrak{g}|^2$ in Eq.~(\ref{eq:cross_section}). At low frequencies, $\mathfrak{g}$ is enhanced for the corotating polarization by superradiance~\cite{Dolan:2008kf}. 

In Fig.~\ref{fig:all_spin_highfreq}, we compare the on-axis Kerr scattering cross sections for short-wavelength ($1.0\leq|\omega|M\leq4.0$) massless bosonic waves ($s=0$, $1$, and $2$). For $s=0$, the cross sections present a glory maximum in the backward direction. For higher-spin waves ($s>0$), the backward-scattered flux is zero in the electromagnetic case ($s = 1$), and negligible in the gravitational case ($s = 2$) above the superradiance threshold of $\omega>am/(2Mr_+)$ with $m=2$.  The angular width of the spiral scattering oscillations of corotating polarizations are wider than counter-rotating ones, again leading to the generation of a net polarization.

\begin{figure*}
	\begin{center}
		\includegraphics[width=0.5\columnwidth]{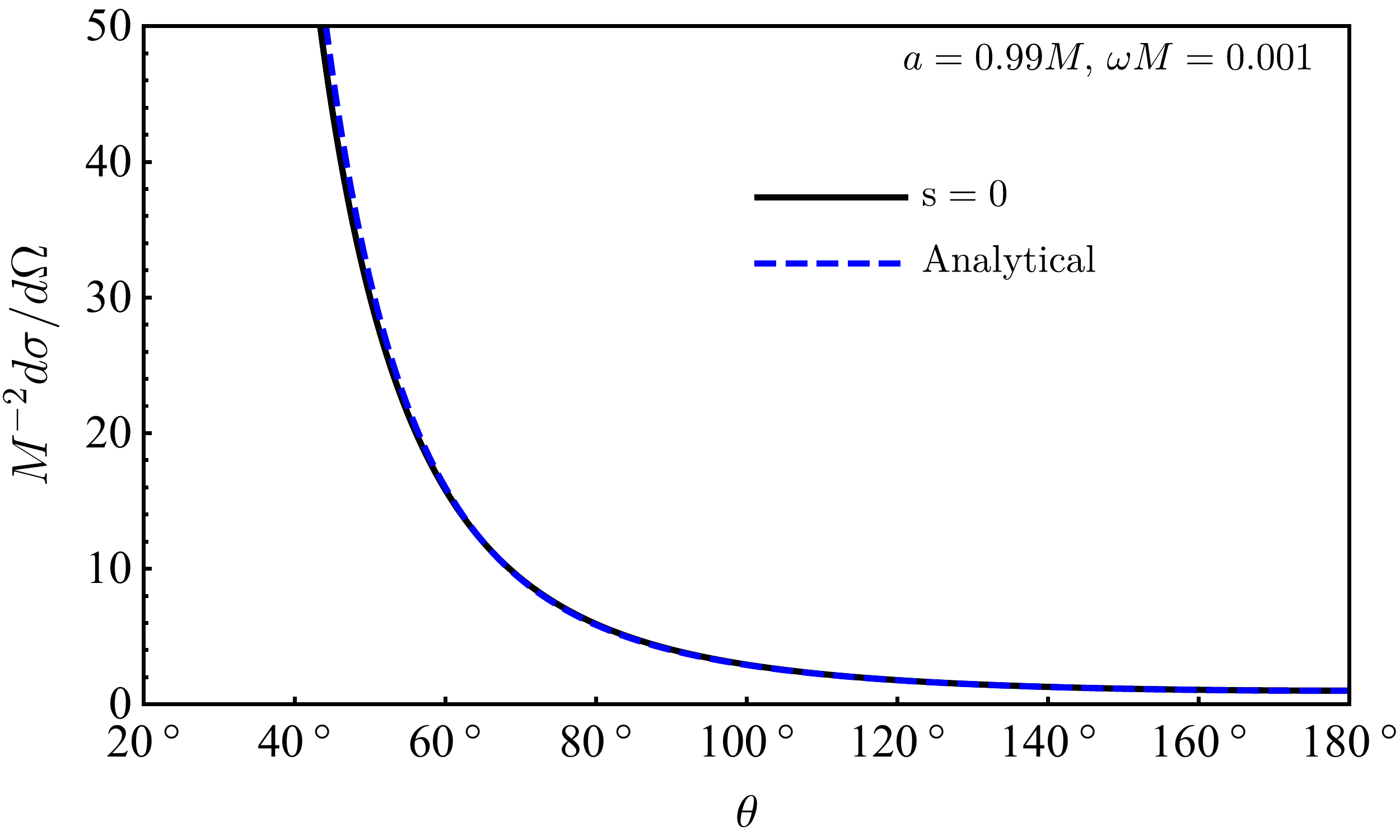}\\
		\includegraphics[width=0.5\columnwidth]{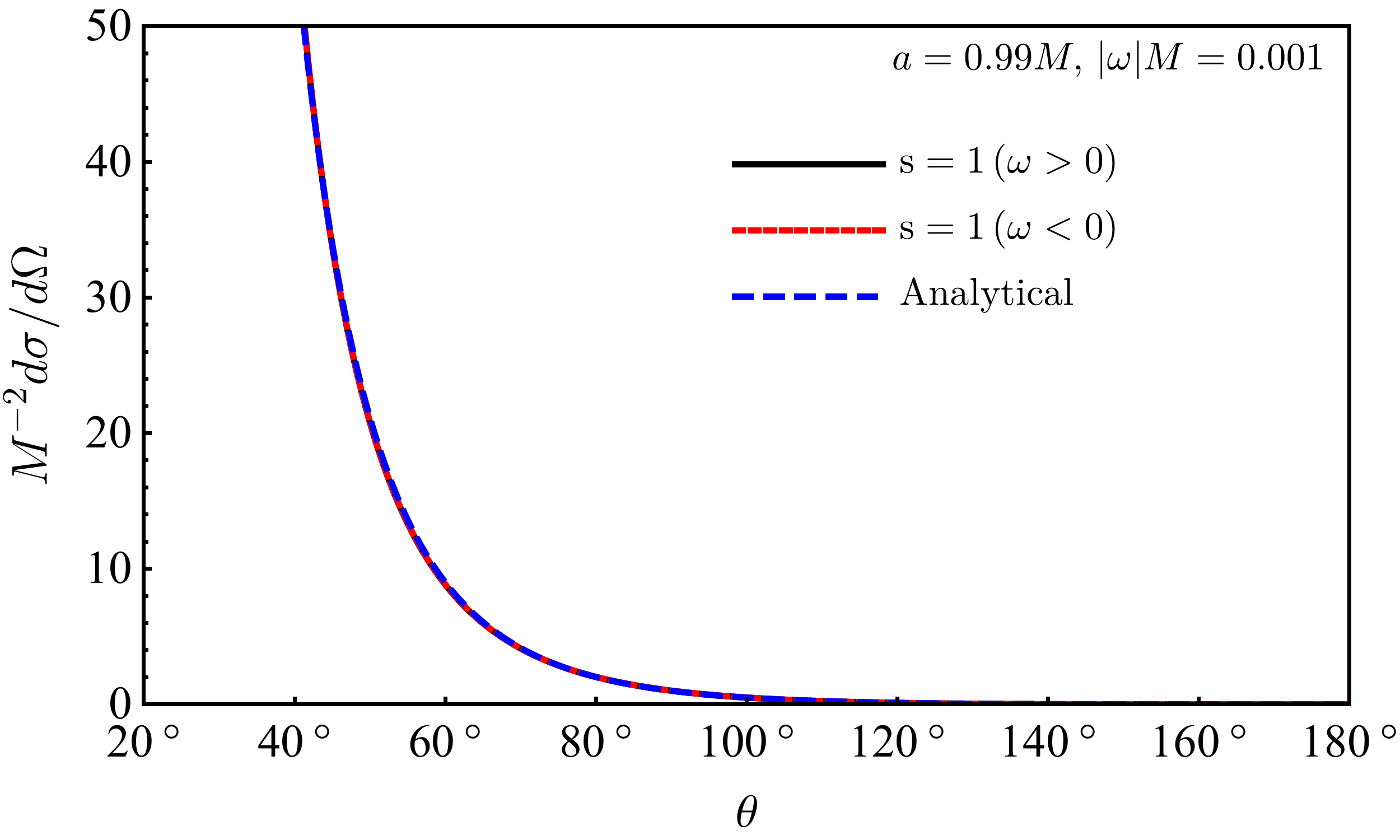}\includegraphics[width=0.5\columnwidth]{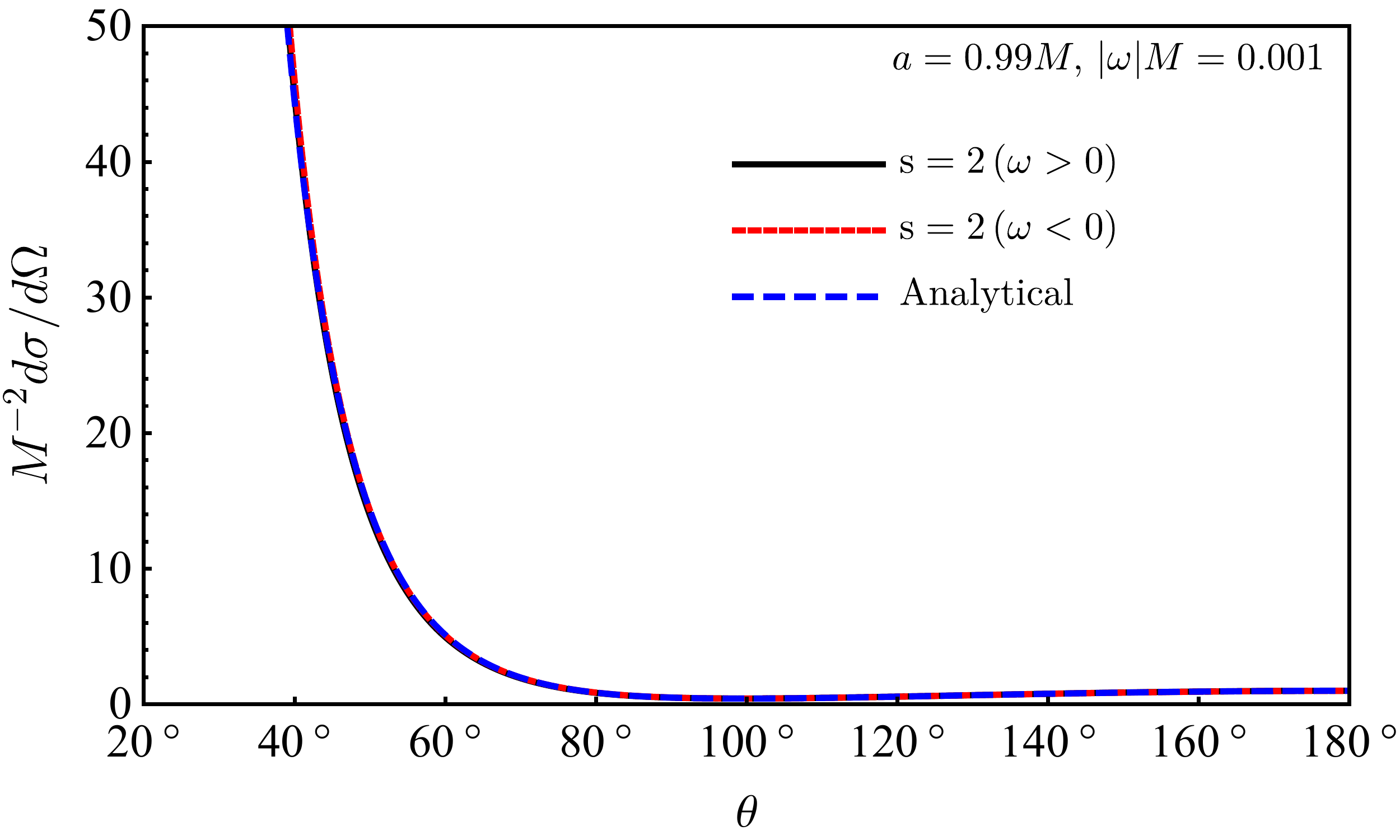}
		\caption{Comparison between numerical and analytical results for the Kerr scattering cross section in the long-wavelength~($|\omega|M=0.001$) regime for the case $a=0.99M$.}
		\label{fig:anal_num}
	\end{center}
\end{figure*}

Finally, in Fig.~\ref{fig:anal_num} we compare the numerical and analytical [computed through Eq.~\eqref{eq:lowfreq_grav_scat}] results for the on-axis Kerr scattering cross section of long-wavelength waves~($|\omega|M=0.001$), in the scalar~(top panel), electromagnetic~(bottom left panel), and gravitational~(bottom right panel) cases. 
We obtain an excellent agreement between the numerical and analytical results.

\section{Concluding remarks}\label{sec:remarks}
We have developed and applied a numerical method to compute the scattering cross section of a Kerr BH for scalar, electromagnetic, and gravitational plane waves, in the special case of on-axis incidence. We found that the scattering cross sections share key features with the static case of spherically symmetric BHs, namely, a forward divergence, spiral scattering oscillations, and a backward glory. A key difference for the Kerr case, however, is that a rotating BH can distinguish between the corotating and counter-rotating circular polarizations of the incident wave \cite{Leite:2017zyb}. This is a subtle effect that is not accounted for in geometrical optics at leading order.

For nonzero spin waves ($s>0$), we have shown that a Kerr BH scatters the two polarizations differently, supporting the interpretation of a coupling between the frame-dragging of spacetime and the field helicity. As pointed out in Ref.~\cite{Dolan:2008kf}, this coupling has the effect of inducing a partial polarization in an initially unpolarized beam. A consequence of this, manifest in e.g.~Figs.~\ref{fig:EM_low} and \ref{fig:EM_high}, is that the net polarization will vary as one varies the scattering angle.

The semiclassical approximation (\ref{eq:glory}) gives a robust estimate of the glory effect, but it does not account for this coupling between helicity and black hole spin. It seems likely that a more refined approximation could be obtained via the complex angular momentum method; see Sec.~IV in Refs.~\cite{Folacci:2019cmc} and \cite{Folacci:2019vtt} for results in the Schwarzschild case. 

Suppose that one could observe a rotating BH illuminated by broadband radiation at a characteristic frequency of $\omega M \sim 1$. While it is not feasible for a solar-system-based observer to significantly change the observing angle $\theta$, it is feasible to observe the system at several wavelengths. Then, the orbiting phenomenon would lead to regular oscillations in the observed flux with wavelength; and moreover, the frame-dragging of the BH would generate regular oscillations in the polarization state, from predominantly left-handed to predominantly right-handed, and back again.

The results presented here are restricted to the special case of a wave that impinges along the direction of the Kerr rotation axis. In the generic off-axis case, the amplitudes (\ref{eq:hel_con}) and (\ref{eq:hel_rev}) are found from a double sum, taken over $j$ and $m$ (azimuthal) modes, which is divergent. In Ref.~\cite{Glampedakis:2001cx}, scalar field ($s=0$) cross sections were calculated by subtracting a Newtonian-type series from this double sum. To treat the higher-spin cases, we propose instead adapting the series reduction method to the off-axis case.

\quad

\appendix

\begin{table}
	\begin{widetext} 
		{
\caption{Numerical data for the scattering cross section $\frac{d\sigma}{d \Omega}$ for the case $a = 0.99M$, $M |\omega| = 2$. The digits in parantheses give an estimate of the numerical error in the final significant figure quoted, found by comparing $n=2$ and $n=3$ iterations of the series reduction method described in Sec.~\ref{sec:series_reduction}, with $\ell_\text{max} = 60$.}
}
\end{widetext}
\begin{tabular}{| r | l | l l | l l |}
\hline
& $s = 0$ &  $s = 1$ & & $s = 2$ & \\
$\theta$ && $a\omega > 0$ & $a \omega < 0 $ & $a\omega > 0$ & $a \omega < 0$ \\
\hline
$20^{\circ}$ & $1.46(3) \times 10^3$ & $1.3(8) \times 10^3$ & $1.39(8) \times 10^3$ & $1.43(8) \times 10^3$ & $1.39(8) \times 10^3$ \\
$30^{\circ}$ & $3.19(1) \times 10^2$ & $3.11(4) \times 10^2$ & $3.00(5) \times 10^2$ & $2.849(8) \times 10^2$ \; & $3.16(2) \times 10^2$ \\
$40^{\circ}$ & $1.191(1) \times 10^2$ & $1.132(3) \times 10^2$ \; &  $97.1(4)$ & $98.6(4)$ & $1.251(5) \times 10^2$  \\
$50^{\circ}$ & $46.99(3)$  & $52.67(2)$ & $51.62(5)$ & $62.3(1)$ & $52.14(7)$ \\
$60^{\circ}$ & $29.102(2)$ & $23.19(5)$ & $31.836(1)$ & $32.05(2)$ & $21.546(8)$ \\
$70^{\circ}$ & $18.829(4)$ & $19.913(2)$ & $14.650(4)$ & $15.163(5)$ & $18.000(2)$ \\
$80^{\circ}$ & $7.234(6)$ & $13.37(3)$ & $13.762(7)$ & $15.518(5)$ & $11.361(2)$ \\
$90^{\circ}$ & $10.181(1)$ & $7.802(3)$ & $9.775(5)$ & $4.378(2)$ & $4.902(4)$ \\
$100^{\circ}$ & $4.9600(3)$ & $10.02(2)$ & $4.5622(4)$ & $9.900(3)$ & $8.884(3)$ \\
$110^{\circ}$ & $4.7089(1)$ & $3.3011(1)$ & $8.1650(8)$ & $1.8505(1)$ & $3.6259(1)$ \\
$120^{\circ}$ & $6.45237(3)$ & $8.427(2)$ & $2.4888(2)$ & $9.1383(3)$ & $4.5424(4)$ \\
$130^{\circ}$ & $1.5715(2)$ & $1.6146(7)$ & $5.5001(1)$ & $1.1628(1)$ & $6.6994(2)$ \\
$140^{\circ}$ & $8.6490(4)$  & $8.86690(4)$ & $4.9005(1)$ & $10.5102(1)$ & $1.4553(1)$ \\
$150^{\circ}$ & $0.95527(3)$  & $0.8830(3)$ & $2.5716(1)$ & $11.7776(1)$ & $12.34590(7)$ \\
$160^{\circ}$ & $12.3040(8)$  &  $16.348(1)$ & $12.5921(6)$ & $1.77444(1)$ & $6.6422(1)$ \\
$170^{\circ}$ & $9.976(1)$  & $5.1772(3)$ & $7.4316(6)$ & $0.014593(2)$ & $0.09270(5)$ \\
$180^{\circ}$ & $77.5562(2)$ & $0$ & $0$ & $1.1515030(2) \times 10^{-3}$ & $2.8316040(5) \times 10^{-4}$ \\
\hline
\end{tabular}
\label{tbl:data}

\end{table}

\begin{acknowledgments}
We thank Tom Stratton for helpful comments and discussions. The authors would like to thank Conselho Nacional de Desenvolvimento Cient\'ifico e Tecnol\'ogico (CNPq) and Coordena\c{c}\~ao de Aperfei\c{c}oamento de Pessoal de N\'ivel Superior (CAPES) -- Finance Code 001, from Brazil, for partial financial support.
This research has also received funding from the European Union's Horizon 2020 research and innovation programme under the H2020-MSCA-RISE-2017 Grant No. FunFiCO-777740.
L.L.~acknowledges the School of Mathematics and Statistics of the University of Sheffield for the kind hospitality while part of this work was undertaken.
S.D.~acknowledges additional financial support from the Science and Technology Facilities Council (STFC) under Grant No.~ST/P000800/1.
The authors thank the anonymous referee for valuable comments and suggestions.
\end{acknowledgments}

\bibliography{refs}

\end{document}